\documentclass[%
reprint,
 amsmath,amssymb,
 aps,
]{revtex4-2}

\usepackage{graphicx}% Include figure files
\usepackage{dcolumn}% Align table columns on decimal point
\usepackage{bm}% bold math
\usepackage[colorlinks]{hyperref}

\usepackage{xcolor}

\newcommand{\rv}{{\bf r}}
\newcommand{\Rv}{{\bf R}}
\newcommand{\pv}{{\bf p}}
\newcommand{\Pv}{{\bf P}}
\newcommand{\nv}{{\bf n}}
\newcommand{\bv}{{\bf b}}
\newcommand{\Nv}{{\bf N}}

\begin{document}

\preprint{APS/123-QED}

\title{Chain trajectories, domain shapes and terminal boundaries in block copolymers}% Force line breaks with \\

\author{Benjamin R. Greenvall}

\affiliation{
Department of Polymer Science and Engineering, University of Massachusetts, Amherst, MA 01003
}

\author{Michael S. Dimitriyev}
\affiliation{
Department of Polymer Science and Engineering, University of Massachusetts, Amherst, MA 01003
}

\author{Gregory M. Grason}
\affiliation{
Department of Polymer Science and Engineering, University of Massachusetts, Amherst, MA 01003
}
 \email{grason@umass.edu}

\date{\today}% It is always \today, today,

\begin{abstract}

The packing geometry of macromolecules in complex mesophases is of key importance to self-organization in synthetic and biological soft materials.  While approximate or heuristic models rely on often-untested assumptions about how flexible molecules ``fit in'' to distinct locations of complex assemblies, physical assemblies derive from ensembles of fluctuating conformations, obscuring the connection between mesophase geometry and the underlying arrangements.  Here, we present an approach to extract and analyze features of molecular packing in diblock block copolymer (BCP) melts, a prototypical soft matter system, based on the statistical description of chain conformations in self-consistent field (SCF) theory.  We show how average BCP chain trajectories in ordered morphologies can be analyzed from the SCF-derived orientational order parameter of chain segments.  We use these extracted trajectories to analyze the features of local packing geometry, including chain bending and tilt, as well as the {\it terminal boundaries} that delineate distinct domains in ordered BCP morphologies.  We illustrate this analysis by focusing on measurable features of packing frustration in 2D (columnar) and 3D (spherical and bicontinuous) morphologies, notably establishing an explicit link between chain conformations in complex morphologies and their {\it medial geometry}.

\end{abstract}

\maketitle

%\tableofcontents

\section{\label{sec:intro}Introduction}

Supramolecular assembly of amphiphilic molecules underlies structure formation in a broad class of material systems, from synthetic surfactants~\cite{Luzzati1967,Scriven1976,Sorenson2011}, liquid crystals~\cite{Chen2020}, and block copolymers~\cite{Leibler1980,Semenov1985_SovPhysJETP,Matsen1994} to intra-cellular assemblies in biology~\cite{Hyde1997_ch4}.  In these systems, molecules spontaneously organize into a set of basic motifs --- spheres, cylinders, layers, networks --- and in high concentration, or neat systems, adopt periodically-ordered arrangements of those motifs.  A generic challenge facing supramolecular assembly, in each of the specific macromolecular contexts, is to understand how molecular degrees of freedom couple into, and ultimately select among, the many possible hierarchical morphologies.  Most conceptual and theoretical frameworks rely on the notion of molecular ``packing” in distinct phases, roughly referring to the set of spatial arrangements of amphiphilic building blocks in a host morphology and its likely thermodynamic costs.  A well-known heuristic associates a tapered, conical shape to amphiphilic units and compares the fit of that local motif into collective packing in competing morphologies (e.g.~spherical vs.~cylindrical micelles)~\cite{Isrealachvili2011}.   In most structurally complex, and often functionally desirable, supramolecular morphologies, packing geometry is expected to be spatially variable, which is a result of frustration between constraints of space-filling at constant density and the presumed thermodynamic preference for uniform local molecular environments~\cite{Anderson1988,Matsen1996, Grason2006,ACShi2021,Duesing1997}.  Examples of these complex phases include so-called bicontinuous, or double-network, phases related to the triply-periodic Gyroid and Diamond minimal surfaces~\cite{Hyde1997_ch4, Schroder-Turk2006}, or complex alloy-like crystals of space-filling micelles, known as the Frank-Kasper phases~\cite{Dorfman2021}.   In these examples, frustration is colloquially associated with molecular packing constraints of filling the nodal junctions of tubular networks and the interstitial regions between sphere-like domains, respectively \cite{Matsen2002,Thomas1987,Grason2006,ACShi2021}.

These scenarios pose a basic and broad question: How do collective configurations of flexible macromolecules ``fit into'' and ``measure'' geometrically complex supramolecular phases?  In this paper, we address this question in the specific context of block copolymer (BCP) melts, based on self-consistent field (SCF) theoretical methods.  While we restrict our analysis to the case of BCP melts, specifically linear diblocks, we consider this system as a prototype for a more general class of macromolecular amphiphiles, most of which exhibit analogs of micellar, columnar, lamellar and bicontinuous mythologies subject to similar packing considerations.  

In general, attempts to connect molecular conformations to complex supramolecular morphologies face several challenges.  Foremost, molecular degrees of freedom are largely ``invisible'' to experimental methods that probe self-assembled morphology.  For example, small-angle scattering as well as electron microscopy methods resolve only spatial patterns of composition --- that is, they resolve spatial ``lumps'' of density of distinct parts of amphiphilic units.  In the context of BCP, this typically amounts to the collective density of different block chemistries, while the underlying chains themselves are not distinguished.  Simulations of either coarse-grained or atomistic models of amphiphiles provide an alternative ``computational microscopy'' on this issue.  Such approaches can be useful for generating direct snapshots of molecular conformations in ordered phases.  Notwithstanding obvious limitations in accurate parametrization of molecular models and computational sampling of sufficiently large time and length scales, such approaches are generally difficult to interpret in terms of direct and spatially-resolved thermodynamic costs, which necessarily depend on ensembles of highly fluctuating conformations.  As noted above, packing models can shed more direct light on the link between molecular geometry and thermodynamics.  In the context of BCP melts, a particularly useful packing model derives from the {\it strong-segregation theory} (SST) of the standard SCF model, and accounts for the local entropic and enthalpic free energies of BCPs by an approximation of microscopic structure based on locally brush-like collections of chains confined within variable wedges that tessellate a space-filling morphology.  A shortcoming of such packing models is that they are based on limited, and largely untested, prior ansatz about packing patterns in a given morphology.  Moreover, the thermodynamic accuracy of these models is limited to certain regimes, e.g.~SST of diblock melts is strictly accurate in the $\chi N \to \infty$ limit, where $\chi$ is the Flory-Huggins parameter, which quantifies repulsion between unlike components and $N$ is the chain length.  Hence, even presuming accurate chain packing ansatz for SST models, the role of finite $\chi N$ fluctuation effects that are relevant to real experimental conditions remains less clear.

\begin{figure}[h!]
\centering
\includegraphics[width=3in]{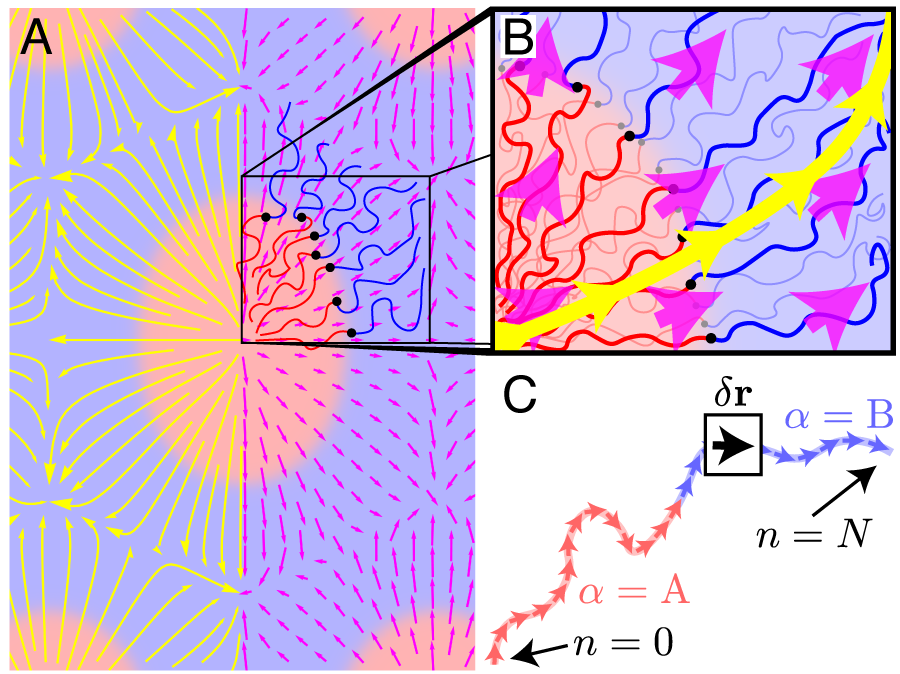}
\caption{\label{fig:1} (A) Depiction of data obtained from SCF calculations, with red and blue regions depicting domains of majority A-block and B-block components, respectively. Magenta arrows show the polar order parameter field $\mathbf{P}(\mathbf{r})$ and yellow curves are stream lines of $\mathbf{P}(\mathbf{r})$. (B) Spatial variations in polar order represent averaged deflections in polymer conformations. (C) Mean polar order arises from microscopic measures of chain flux, the ensemble average over all chain conformations (such as the one depicted) of the orientation $\delta \mathbf{r}$ joining segment $n$ to segment $n+1$ along a chain oriented with respect to ends at $n=0$ and $n=N$.}
\end{figure}

Numerical implementations of the SCF model of BCP are arguably a ``gold standard'' for modeling equilibrium morphologies at finite segregation, at least sufficient far from the critical point (typically for $\chi N \gtrsim 40$) where composition fluctuations play a small role.  This approach provides a fully statistical description of chain fluctuations in competing morphologies without any \textit{a priori} assumptions on the molecular packing.  However, while the SCF theory is built upon the statistics of BCP chain conformations, traditional SCF implementations, like in the case of current experimental methods, are cast only in terms of the scalar composition fields of block components, leaving the locations and arrangements of underlying chains unresolved.

In this article we present and illustrate an approach to map the geometry of chain conformations in BCP melts based on the SCF theory.  We exploit the fact that ordered solutions of SCF, even in the standard Gaussian chain model, are described by orientational order parameters describing local chain ``trajectories'' in the structure.  We show how these mean trajectories can be computed from finite $\chi N$ SCF solutions of ordered phases, and argue that they extract the key ``chain packing'' degrees of freedom from an ensemble of fluctuating chain conformations in a spatially resolved manner.  We apply this approach to consider distinct motifs of chain packing in complex morphologies and compare to prior heuristic notions of frustration in micellar and bicontinuous network phases, particularly in the large $\chi N$ regime.  We show that trajectories extracted from SCF calculations can be used to analyze specific geometric features of the morphology, including the tilting and kinking at the intermaterial dividing surface (IMDS) as well as the so-called {\it terminal boundaries} that represent the contacting ``ends'' of brush-like domains.  We exploit this approach to analyze how these geometric signatures of chain packing vary with structural features of the morphology as well as physical parameters of the diblocks themselves, including composition and conformational asymmetry.

The remainder of this article is organized as follows.  In Sec.~\ref{sec:methods} we present our method of extracting chain trajectories from SCF solutions of diblock melts, as well as what we call the {\it association map} that relates spatial regions in the solution to a particular point on the IMDS.  In Sec.~\ref{sec:2D} we apply these methods to analyze the variation of tilting and bending of chains at the IMDS in columnar phases of different symmetries.  We consider the shapes of terminal boundaries in the packing as a function of the anisotropy of the columnar domain cross-section.  In Sec.~\ref{sec:3D}, we turn to three dimensional frustrated morphologies, illustrating and analyzing chain packing in a complex Frank-Kasper (A15) phase as well as a bicontinuous (double-gyroid) phase.  This latter analysis provides direct evidence from a fluctuating chain description of a recently proposed ``medial packing'' picture in complex BCP assemblies.

\begin{figure}[h!]
\centering
\includegraphics[width=3in]{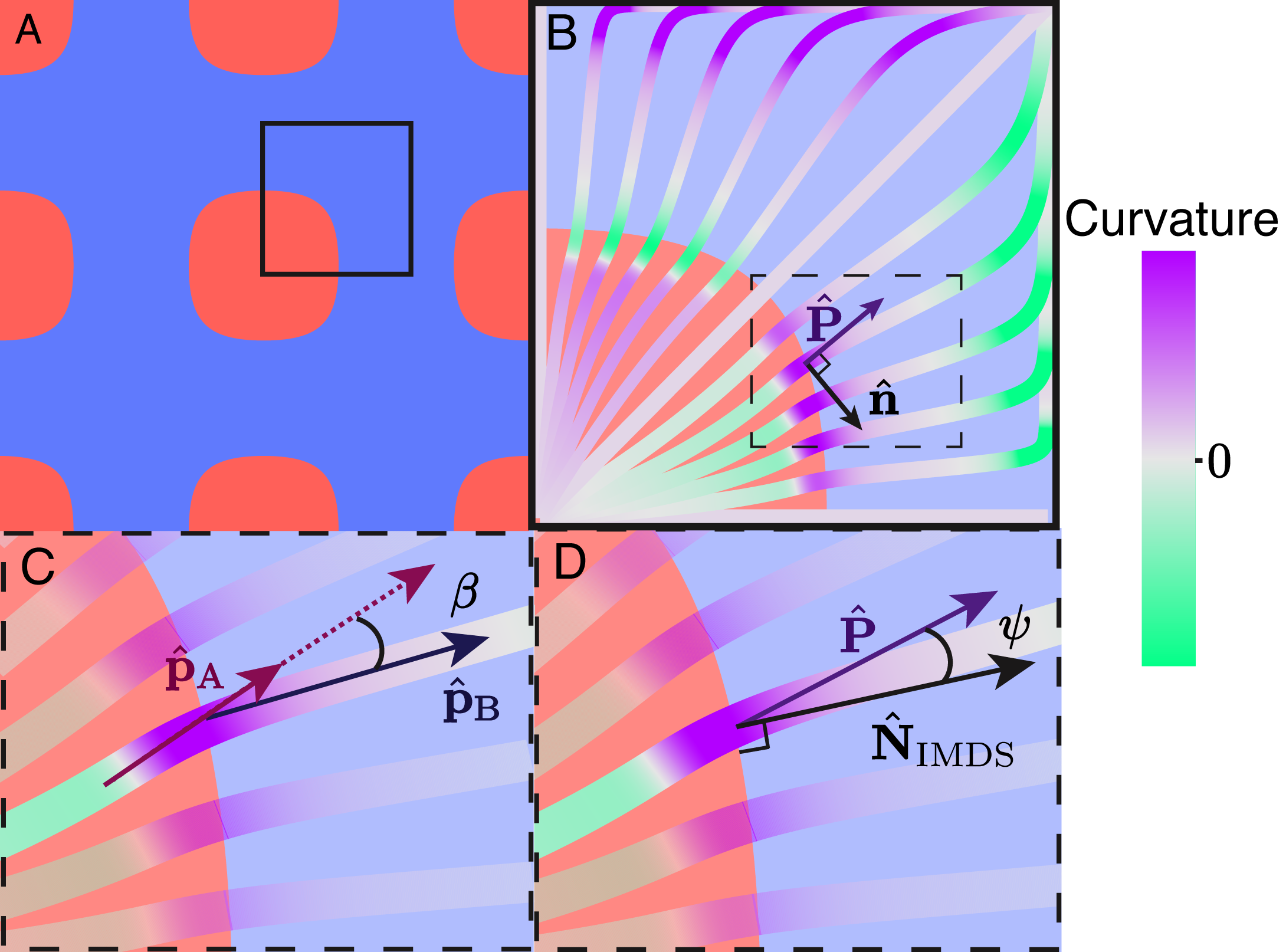}
\caption{\label{fig:local_structure} Relationships between trajectories and domain morphology. (A) Depiction of $p4mm$ columnar phase ($f=0.3$, $\chi N = 100$, $\epsilon = 5.0$) with slightly faceted IMDS. (B) Distribution of signed curvature $\kappa$ (here, $\kappa > 0$ is left-handed and $\kappa < 0$ is right-handed) for a variety of trajectories, showing maxima and minima near the IMDS and near the outer terminal boundary. (C) Measure of chain kinking, given by the angle $\beta$ between $\hat{\mathbf{p}}_{\rm A}$ and $\hat{\mathbf{p}}_{\rm B}$, evaluated at the IMDS. (D) Measure of chain tilt, given by the angle $\psi$ between $\hat{\mathbf{P}}$ and the local IMDS normal $\hat{\mathbf{N}}_{\rm IMDS}$.}
\end{figure}

\section{\label{sec:methods}Methods}

Here, we outline an approach to reconstruct chain trajectories and packing geometry from numerical solutions of the SCF equations for BCP melts.  In this article, we illustrate the approach for linear AB diblock copolymers, although the approach may be generalized to other architectures and multi-chain mixtures.  We consider chains of $N$ total segments, where $N_{\rm A} = f N$ ( $N_{\rm B} = (1-f)N$) are A-type (B-type).  Segments are defined to have equal volume $\rho_0^{-1}$, but may have unequal statistical lengths, $a_{\rm A}$ and $a_{\rm B}$, corresponding to a {\it conformational asymmetry} $\epsilon=a_{\rm A}/a_{\rm B}$.  Our approach applies to the ``standard'' Gaussian chain model SCF of melts~\cite{Matsen2002}, where interactions between A and B-type segments are parameterized by the Flory-Huggins parameter, $\chi$.  The analysis that follows relies on mean-field solutions to the SCF equations for the chain-end distribution functions, as is achieved through several well-known approaches~\cite{Matsen2002, Matsen2005, Fredrickson2002}, although results in the present article are derived from the Polymer Self-Consistent Field (PSCF) code~\cite{Arora2016} (\url{https://pscf.cems.umn.edu/}).  Supporting codes for extracting (polar) orientational order parameters (discussed in Appendix~\ref{app:op}), reconstructing trajectories and analyzing packing geometry of PSCF solutions are provided (\url{https://doi.org/10.7275/1b2p-q547}).

% 2 parts:
% 1. define trajectories 
% 2. association map

\subsection{Chain Trajectories and local packing geometry}
\label{sec: trajectories}

Our approach to reconstruct chain trajectories, shown schematically in Fig.~\ref{fig:1}, is based on the (mean field) polar orientational order parameter computed from SCF introduced in ref.~\cite{Prasad2017}.  This parameter derives from chain end distribution functions $q_{\pm} ({\bf r}, n)$, which describe the statistical weights of chain conformations that ``diffuse'' from their free ends at $n=0$ (+) and $n=N$ (-) to the $n^{\rm th}$ segment at position ${\bf r}$ in the melt.  The probability that the $n^{\rm th}$ segment of a chain in the melt is at ${\bf r}$ is proportional to the joint probability that both ends reach this point $q_+ ({\bf r},n)q_- ({\bf r},n)$, such that the mean-field local volume fractions of the $n^{\rm th}$ segment at $\rv$ is
\begin{equation}
\varphi(\rv, n) = \frac{\rho_0^{-1}}{{\cal Q}}q_+ ({\bf r},n)q_- ({\bf r},n)
\end{equation}
and the composition fields (i.e.~scalar order parameters) of $\alpha =$ A or B segments are 
\begin{equation}
\phi_{\alpha} ({\bf r}) =  \int_{n \in \alpha} {\rm d}n ~ \varphi(\rv, n),
\end{equation}
where ${\cal Q} = V^{-1} \int {\rm d}^3 {\rm r}~ q_+ ({\bf r},n)q_- ({\bf r},n)$ is the normalized single-chain partition function for a total volume $V$.  

To model the {\it trajectories}, we consider the orientational distribution of chain steps from the $n^{\rm th}$ to the $(n+1)^{\rm th}$ segments, described by the vector $\delta {\bf r}$, that is oriented from the $n=0$ (A-block) end toward the $n=N$ (B-block) end, as shown schematically in Fig.~\ref{fig:1}C.  
%Specifically, $\langle \delta {\bf r}  \rangle/a$, the mean orientation of random-walk steps at point ${\bf r}$ from $n$ to $n+1$ is proportional to {\it chain flux} operator
The mean orientation $\langle \delta {\bf r}  \rangle/a$ of random-walk steps at point ${\bf r}$ from $n$ to $n+1$ is proportional to {\it chain flux} operator
\begin{equation}
{\bf J} ({\bf r}, n) =  \frac{\rho_0^{-1}}{6 {\cal Q}} \Big[q_+({\bf r}, n) \nabla q_-({\bf r}, n) -q_-({\bf r}, n) \nabla q_+({\bf r}, n)\Big] \, .
%    \frac{\rho_0^{-1}}{{\cal Q}} \int d^3 (\delta {\bf r}) ~ q_+({\bf r}- \frac{\delta {\bf r}{2},n)  q_-({\bf r}+ \frac{\delta {\bf r}{2},n+1) p(\delta {\bf r} ) \frac{\delta {\bf r}}{a_\alpha} \\
\end{equation}
Specifically, the relation
\begin{equation}
\label{eq: fluxor}
{\bf J} ({\bf r}, n)= \Big\langle \frac{\delta {\bf r} }{a_\alpha }\Big\rangle_{({\bf r},n)} \varphi(\rv, n)
\end{equation}
follows from the average of $ \delta {\bf r} $ weighted by the chain-end probabilities $ q_+({\bf r}- \frac{\delta {\bf r}}{2},n)  q_-({\bf r} + \frac{\delta {\bf r}}{2},n+1)$ times probability of a random-walk step from ${\bf r}- \frac{\delta {\bf r}}{2}$ to ${\bf r}+ \frac{\delta {\bf r}}{2}$.  Notably, the same differential form follows from both the ``Gaussian thread'' model as well as the continuum ($N\gg1$) limit of a freely jointed chain.  

Given the relation in Eq.~(\ref{eq: fluxor}), it is straightforward to construct the {\it mean paths} of chains where the $n_0^{\rm th}$ segment passes through $\rv_0$, described by the function $\Rv_{(\rv_0,n_0)}(n)$ by identifying the path tangent $\partial_n \Rv_{(\rv_0,n_0)}$ as proportional to the local chain flux, i.e.,
\begin{equation}
\partial_n \Rv_{(\rv_0,n_0)} = {\bf J} \big(\Rv_{(\rv_0,n_0)}(n), n\big)/\varphi\big(\Rv_{(\rv_0,n_0)}(n), n \big) ,
\end{equation}
which can be integrated subject to the initial condition $\Rv_{(\rv_0,n_0)}(n_0) = \rv_0$.  In equilibrium states of BCP melts, a given point is intersected by an ensemble of chain paths, leading to a distribution of segment numbers (i.e.~a distribution of $n$) at a given point.  As our interest is in the statistical average of conformations at distinct spatial points, we consider the average over all chain conformations with segments at a given point, information that is encoded in the {\it polar order parameters}
\begin{equation}
\pv_\alpha (\rv)  = \int_{n \in \alpha} {\rm d}n ~ {\bf J} ({\bf r}, n) ,
\end{equation}
which give the local ``flux'' of trajectories of all $\alpha$-type segments at a point $\rv$.  

We define the mean {\it trajectories} of chains in terms of the total polar order parameter $\mathbf{P}$, the sum of averages over local densities of both segment types,
\begin{equation}
 \Pv (\rv) = \pv_{\rm A} (\rv)+ \pv_{\rm B} (\rv) .
\end{equation}
In effect, mean trajectories are simply the {\it integral curves} of the vector field $\Pv (\rv)$.  Defining $\Rv_{\rv_0} (t)$ as trajectory that passes through point $\rv_0$ at $t=0$, the ``flow'' of trajectories along $\Pv(\rv)$ satisfies
\begin{equation}
\label{eq:int_curves}
\partial_t \Rv_{\rv_0} (t) = \Pv \big(\Rv_{\rv_0} (t)\big) ,
\end{equation}
subject to the initial condition $\Rv_{\rv_0} (0)=\rv_0$.  Note that $t$, which parameterizes the flow along a trajectory from the A- to B-end of chains, has no specific relation to the distance between segments along the paths.  An example of the relationship between the polar order parameter (magenta vectors) and reconstructed trajectories (yellow stream lines) is shown in Fig.~\ref{fig:1}A-B.

In what follows, we analyze the geometry of chain trajectories as illustrated schematically in Fig.~\ref{fig:local_structure}.  First we can analyze the {\it bend} of trajectories $\bv(\rv)$ from the unit vector of polar orientation $\hat{\pv}(\rv)$,
\begin{equation}
   \bv(\rv) = (\hat{\pv} \cdot \nabla)  \hat{\pv}  \equiv \kappa(\rv) \hat{\nv}(\rv) ,
\end{equation}
where $\kappa(\rv)$ and $\hat{\nv}(\rv)$ are the curvature and normal to the trajectory at $\rv$.  

As shown in the example of Fig.~\ref{fig:local_structure}B, trajectories are largely straight, with the exception of two regions.  First are the portions of trajectories near the \textit{outer terminal boundaries}, where trajectories from one domain meet trajectories flowing in from another domain/region. In general, this leads to localized bending of trajectory orientation parallel to those boundaries.  We show in Appendix~\ref{app:dithering}, however, that such ``high deflections'' in the distal ends of trajectories correspond to overlap between opposing brushes where the chain loses orientation, corresponding to a regions where $|\pv(\rv)|\to 0$.  Hence, for the purposes of focusing on the strong-segregation features of chain packing, deflections in this distal zone can be neglected.

Additionally, some chain configuration show localized bend at the IMDS, which is defined at the points where $\phi_{\rm A} (\rv)=\phi_{\rm B} (\rv)= 1/2$. These sharp bends, or {\it kinks}, are anticipated in certain SST models of BCP melts as one means to negotiate the conflicting demands of chain packing~\cite{Xi1996,MatsenBates1995,Olmsted1998}.  We analyze the kink angle $\beta$ from SCF solutions, which we take to be the difference between the polar orientation on the A and B side of the IMDS.  Since the polar order parameter transforms from all A-type to B-type segments over the interfacial width, in practice it is most convenient to do this by comparing the values of $\hat{\pv}_{\rm A}$ and $\hat{\pv}_{\rm B}$ at the IMDS, or
\begin{equation}\label{eq:kink_angle_cos}
   \cos \beta = \hat{\pv}_{\rm A} (\rv) \cdot \hat{\pv}_{\rm B} (\rv) , \ \ \ \ \ {\rm for} \ \rv \in {\rm IMDS} .
\end{equation}
In Appendix~\ref{app:bend_threshold}, we compare this measure of kink to the angle between orientations $\hat{\Pv}$ along the same trajectory, but at points just ``up-/down-stream'' of the composition gradient at the IMDS, and find that both measures capture at least the same qualitative features of the packing and its dependence on BCP parameters.  

A related feature of local chain packing geometry is the {\it tilt} of chains relative to the IMDS.  While the simplest models of packing assume that the mean trajectories of chains extend {\it normal} relative to the IMDS, such a pattern may come into conflict with constraints of filling space at constant density.  This feature of smectic-$C$-like packing is well appreciated in packing models of lyotropic phases of amphiphiles~\cite{Hamm2000, Chen2017}, particularly in complex, bicontinuous phases.  More recently, a SST model of network phases based on the so-called {\it medial packing} suggested that tilt is a generic feature of BCP melt packing as well~\cite{Reddy2022,Dimitriyev2023}.  To assess the degree of tilt, we measure the angle $\psi$ between the mean chain orientation $\hat{\mathbf{P}}$ and the IMDS normal $\hat{\Nv}_{\rm IMDS}$, 
\begin{equation}
\cos \psi = \hat{\Pv} (\rv) \cdot \hat{\Nv}_{\rm IMDS}(\rv)  \ \ \ \ \ {\rm for} \ \rv \in {\rm IMDS} ,
\end{equation}
where $\hat{\Nv}_{\rm IMDS} \equiv - \nabla \phi_{\rm A} / |\nabla \phi_{\rm A}|$ defines the normal to the IMDS, where $\phi_{\rm A} (\rv) = 1/2$.

\subsection{Association map, domains and terminal boundaries}

\begin{figure}
\centering
\includegraphics[width=2.75in]{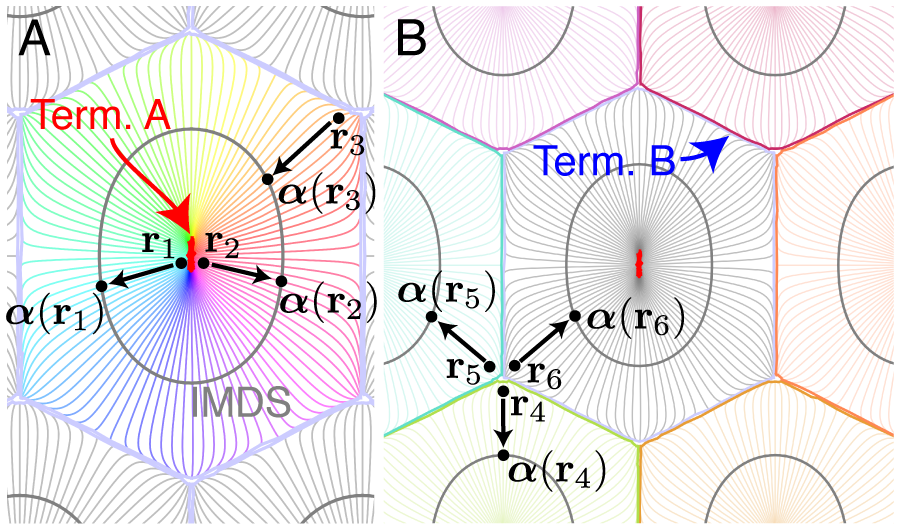}
\caption{\label{fig:global_structure} Global structure of the association map and terminal boundaries in a columnar morphology. (A) focuses on a domain, with the inner terminal boundary (``Term.~A'') highlighed in red, as well as example points $\{\mathbf{r}_1,\mathbf{r}_2,\mathbf{r}_3\}$ with their associations $\{\bm{\alpha}(\mathbf{r}_1),\bm{\alpha}(\mathbf{r}_2),\bm{\alpha}(\mathbf{r}_3)\}$ onto a single IMDS. (B) shows the association maps onto distinct IMDSs with separate outer terminal boundaries (``Term.~B'') of distinct domains highlighted in different colors. Example points $\{\mathbf{r}_4,\mathbf{r}_5,\mathbf{r}_6\}$ associate to distinct IMDSs, despite their relatively close proximity in space.}
\end{figure}

Beyond the local geometry of chain trajectories, packing is generally concerned with the shapes of {\it domains} and the distributions of conformations of the chains that compose those domains.  Following ref.~\cite{Reddy2021}, we define a {\it domain} as the set of points in space occupied by BCP chains whose junctions belong to a particular IMDS, a property we denote as {\it association}.  Here, we construct association in terms of the mean trajectories extracted from SCF calculations.  Since every point on the IMDS is associated with a distinct trajectory, all points in a domain (except at terminal boundaries) can be mapped to single, corresponding points on an IMDS; this defines the \emph{association map}.

The association map $\bm{\alpha}(\mathbf{r})$ is constructed by selecting an arbitrary point $\mathbf{r}$ and then propagating from that point using Eq.~(\ref{eq:int_curves}) until the IMDS is reached.  In terms of the notation defined above,
\begin{equation}\begin{split}
 &\bm{\alpha}(\mathbf{r}) = \Rv_{\rv} (t_{\rm IMDS}),\\
 &\mkern+32mu {\rm such\, that}\,\, \phi_{\rm A} \big(\Rv_{\rv} (t_{\rm IMDS}) \big) = 1/2\, .
\end{split}\end{equation}
Note that this introduces a map from the points on the IMDS to all points within a domain, corresponding to the trajectories that pass through the the IMDS.  

As shown in Fig.~\ref{fig:global_structure}A, the preimage of the association map (i.e., $\bm{\alpha}^{-1}$), consists of trajectories passing through the IMDS that extend through the A and B portion of the domain.  The ends of these trajectories mark the {\it terminal boundaries} of the domains.   The association maps $\bm{\alpha}$ can, however, be multi-valued for a subset of points, namely those that lie on the terminal boundaries of a domain.  This derives from the fact that at the contact point between two locally opposing brush regions, the segments at that point are equally likely to be anchored (i.e.~their junctions are located) at different IMDS points. 
%The location of the terminal boundaries can be determined by the non-invertability of the association map, since these points are not associated with a unique point on the IMDS.  % (the association map is not really invertible anywhere
We thus determine the locations of the terminal boundaries by searching for regions where the association map fails to be continuous.
Operationally, we search for regions where the Jacobian matrix,
\begin{equation}
    \Lambda_{ij} (\rv) = \frac{\partial \alpha_i}{\partial r_j}
\end{equation}
becomes singular, i.e.~where numerically, the principal eigenvalue $\Lambda$ of the matrix $\bm{\Lambda}$ tends to diverge ($|\Lambda| \to \infty$).

As outlined in Appendix~\ref{app:term}, to implement this numerically from SCF solutions, we remesh our solutions with triangular (2D) or tetrahedral (3D) elements, and interpolate $\alpha(\rv)$ onto vertices.  This remeshing allows us to increase resolution near to singular points in the association map as needed.  The discrete approximation of $\bm{\Lambda}$ can be computed using these finite elements, by noting that under the action of the association map $\bm{\alpha}$, each element is stretched and compressed in different directions, characterized by the affine matrix $\Lambda_{ij}$.  We search for regions where this deformation diverges in terms of the maximal stretch of this transformation.  The maximal stretch is  determined by the maximal eigenvalue of the matrix product $\bm{\Lambda}$; we can then identify terminal boundaries as passing through facets whose maximal eigenvalue is larger than some threshold value $\Lambda_{\rm thresh}$.  As noted in Appendix~\ref{app:term}, the maximal distortion is intrinsically limited by mesh resolution, and in practice some analysis of the distribution of element distortions is used to numerically delineate the singular from the non-singular regions.  Finally, in order to generate meshes of these terminal boundaries, we start with a mesh of the IMDS (i.e.~the isocontour $\phi_{\rm A} (\rv) = 1/2$) and flow vertices along trajectories until threshold stretch eigenvalue is reached.

Examples of the terminal boundaries for a 2D columnar morphology are highlighted in Fig.~\ref{fig:global_structure}.  The terminal boundary of the A region (highlighted as red in Fig.~\ref{fig:global_structure}A) corresponds to points where chains are associating to distant points on the {\it same} convex IMDS, which we call an {\it inner terminal boundary}.  As we describe below, the fact that this inner terminal boundary is in general not point-like indicates that the chains are not focusing to the centroid of the convex domain, but instead their terminal ends spread over a finite 1D region of the cross-section.  The terminal boundaries of the B regions (highlighted as different colors according to distinct domains in Fig.~\ref{fig:global_structure}B) corresponds to points where chains are associating to {\it distinct} IMDSs.  Denoting these as {\it outer terminal boundaries}, they clearly split the melt into distinct domains and function similar to Voronoi (or Wigner Seitz) cells of the assembly.  Crucially, unlike Voronoi cells, which are defined in terms of distances from central points, terminal boundaries are defined by the actual underlying molecular conformations.  In other words, the geometric features of these terminal boundaries are {\it selected by the molecules} as means to minimize the system's free energy. As we show below, this distinction means that shapes of terminal boundaries vary with BCP parameters controlling segregation, composition and chain stiffness, unlike Voronoi cells, which are fixed for a given space group.

\begin{figure*}[t]
    \centering
    \includegraphics[width=0.8\textwidth]{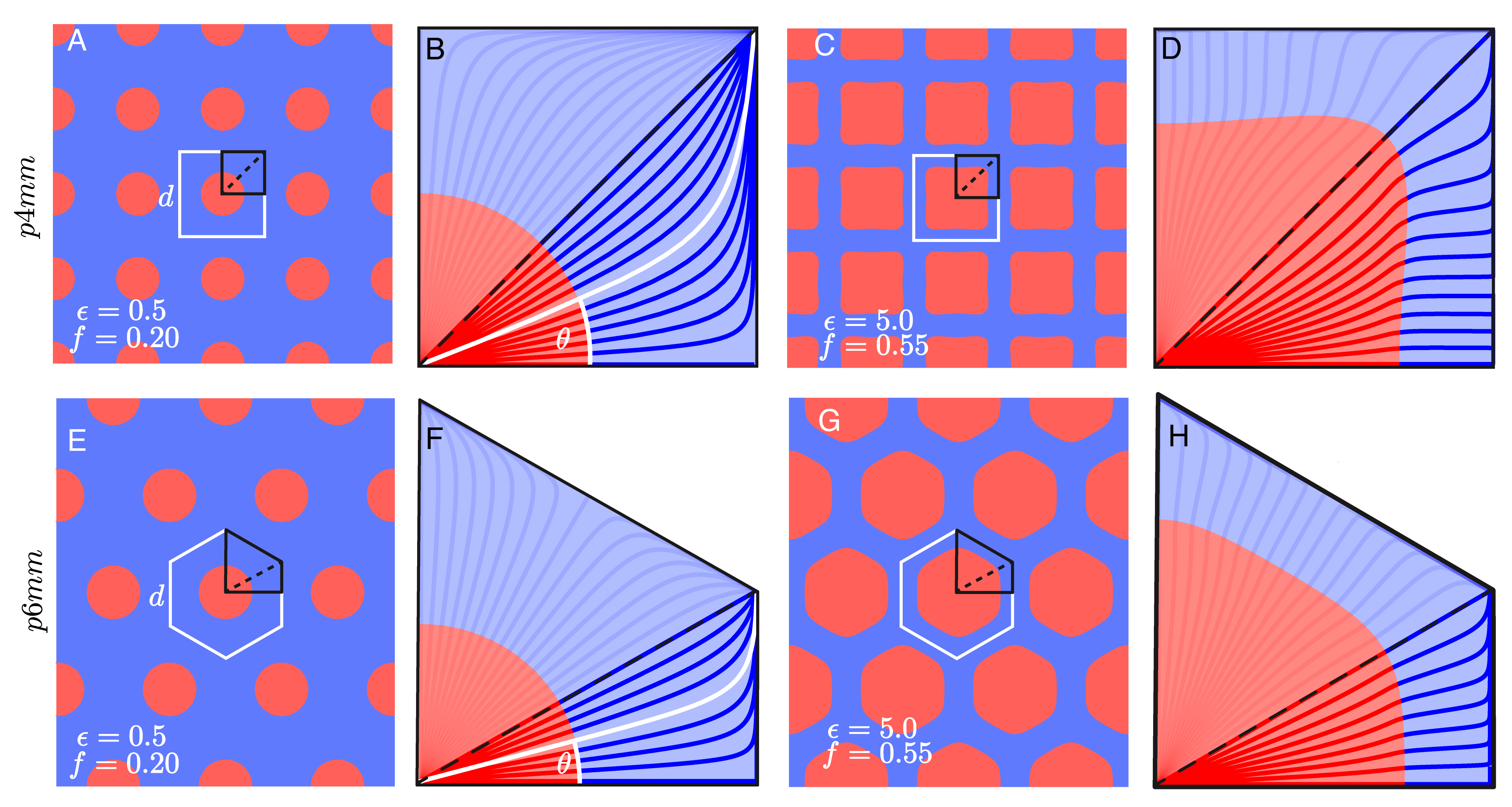}
    \caption{\label{fig:traj_wedges} Domain structure of square cylinders ($p4mm$) with cell edge length $d$ at fixed $\chi N = 100$ and for parameters (A) $f=0.20$, $\epsilon = 0.5$ and (C) $f=0.55$, $\epsilon = 5.0$, with corresponding trajectories in (B) and (D), respectively. Selected trajectories pass through the IMDS at angle $\theta$ with respect to the $x$-axis. The fundamental domain $0 \leq \theta < 45 ^\circ$ is highlighted. Domain structure of hexagonal cylinders ($p6mm$) with cell edge length $d$ at fixed $\chi N = 100$ and for parameters (E) $f=0.20$, $\epsilon = 0.5$ and (G) $f=0.55$, $\epsilon = 5.0$, with corresponding trajectories in (F) and (H), respectively.  The fundamental domain $0 \leq \theta < 30 ^\circ$ is highlighted. Chosen parameters approximately follow the $p6mm$-lamella equilibrium phase boundary as reported in ~\cite{Grason2004}}
\end{figure*}

\section{\label{sec:2D}Columnar morphologies}

We first illustrate the analysis of chain packing geometry and the terminal boundaries by considering 2D columnar mesophases.
Columnar phases have long provided a testing ground for ideas related to packing frustration and how it alters domain morphology and thermodynamics in BCP melts~\cite{Helfand1980, Fredrickson1993, Olmsted1994, Olmsted1998, Matsen1996, Grason2004, Grason2006}.  The conflict between a thermodynamically uniform favored cylindrical geometry and then need to ``fill the empty corner'' that would be created by closed packing of cylinders is generally considered to be a source of frustration~\cite{ACShi2021}, which is resolved by some combination of variable IMDS curvature and chain trajectory deflection towards the interstices in the packing.  In SST models, Olmsted and Milner  formulated two variants of chain trajectories that satisfy packing constraints~\cite{Olmsted1994, Olmsted1998}.  On one hand, if the IMDS remains perfectly round, chains can kink from radial orientation towards the interstices to satisfy local volume constraints.  On the other hand, if the IMDS perfectly copies the shapes of the (outer terminal boundary) unit cell, chains can retain straight trajectories, but will clearly incline, or tilt, with respect to the IMDS.  Known respectively as the {\it kinked}- and {\it straight}-path ansatzes, these represent two extremes for how BCP chains resolve packing frustration, and of course, variants that interpolate between these extremes suggest that SST packing~\cite{Grason2004, Grason2006}, even in columnar morphologies where frustration is relatively weaker than other morphologies, can vary significantly with chain parameters.  For what follows we consider variable A-block volume fraction $f$ as well as variable conformational asymmetry $\epsilon$, which controls the elastic asymmetry between A- and B-brush regions.

In the context of these prior motifs for chain packing in columnar morphologies, we explore the chain trajectories in finite, but generally large, values of $\chi N$, based on SCF solutions which impose no assumptions of the packing motif and therefore reflect at least some degree of conformational fluctuations absent from SST.  We first consider the classical hexagonal cylinder phase, as well as the more frustrated square phase to illustrate how inter-domain packing alters the subdomain geometry of chain packing.  Next, we consider a lower symmetry family of cylinder packings to explore the link between domain anisotropy, terminal boundary geometry, and chain packing.

\begin{figure*}[t]
\centering
\includegraphics[width=\textwidth]{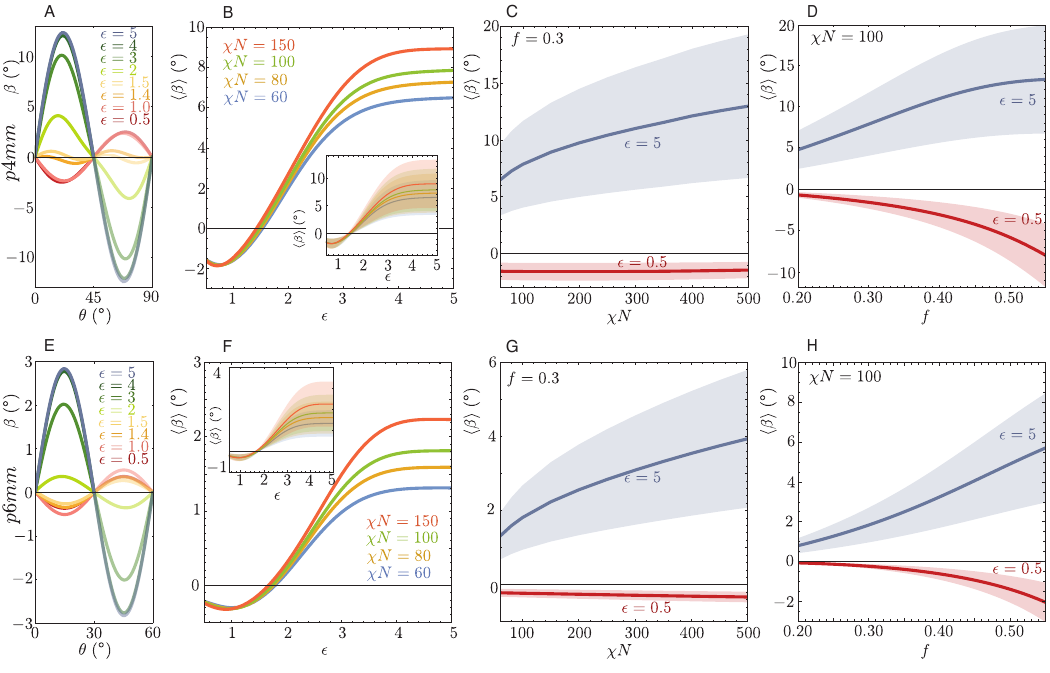}
\caption{\label{fig:bend_horizontal} (A) Kink angle as a function of polar angle $0 \leq \theta \leq 90 ^\circ$ for the square cylinder phase at $\epsilon = 0.5 - 5.0$ (fixed $\chi N = 100$, $f=0.3$). Intermediate values of $\epsilon \sim 1.4-1.5$ show an inflection of the kink angle within the fundamental domain $0 \leq \theta \leq 45 ^\circ$. (B) Average of the kink angle, $\langle \beta \rangle$, taken over a fundamental domain as a function of $\epsilon$ for $\chi N = 60-150$ (fixed $f=0.3$) showing that the kink angle tends towards negative values for low $\epsilon$, positive values for high $\epsilon$ and passes through 0 at $\epsilon \simeq 1.45 \pm 0.0.5$. Inset shows the variation in the kink angle, where the shaded region extends for a single standard deviation on either side of the mean. Finally, bending angle statistics for extreme values of conformational asymmetry ($\epsilon = 0.5$ and $5.0$) are shown as a function of (C) $\chi N$ and (D) $f$, confirming the consistency of the sign of the kink angle. (E)-(H) show these data for the hexagonal cylinder phase ($p6mm$), where, notably, the kink inversion occurs at slightly higher $\epsilon$, $\epsilon \simeq 1.75 \pm 0.05$.}
\end{figure*}

\subsection{\label{sec:2D_p4p6} Trajectories near the IMDS: Packing in Hexagonal ($p6mm$) and Square ($p4mm$) Cylinders}

We analyze the chain trajectories of square and hexagonal lattice columnar phases extracted from SCF solutions for a range of chain parameters.  Examples in Fig.~\ref{fig:traj_wedges} highlight trajectories within the fundamental domain (or, asymmetric wedge) of each morphology, at variable $f$ and $\epsilon$ with fixed $\chi N = 100$.

Of the ordered columnar phases, the hexagonal cylinder (with space group $p6mm$, shown in Fig.~\ref{fig:traj_wedges}E) is most generically an equilibrium phase for linear diblocks.  Heuristically, this is often attributed to the fact that hexagonally-closed packing requires the lowest void density ($<10\%$), so that distortions of chain packing to fill the gaps away from cylindrical symmetry is minimal compared to other packings~\cite{Thomas1986, ACShi2021, Grason2006}.  Notably, the square columnar phase (space group $p4mm$, shown in Fig.~\ref{fig:traj_wedges}A) has been observed in some block copolymer architectures~\cite{Zheng1995,Dorin2014, Aissou2023}, as well as under template-directed assembly conditions~\cite{Tang2008}, and is a morphology that is expected to be relatively challenging for chains to occupy, requiring larger deflections from radial trajectories.  We quantify these distinctions by comparing  trajectories extracted from SCF solutions of square and cylinder packing.

The reconstructed chain trajectories in Fig.~\ref{fig:traj_wedges}B ($p4mm$) and Fig.~\ref{fig:traj_wedges}F ($p6mm$) clearly illustrate the frustrated nature of packing in columnar phases, where trajectories bend away from the direction of the nearest cylindrical neighbor ($\theta = 0$) towards the next nearest neighbor ($\theta = 45 ^\circ$ or $30 ^\circ$ for $p4mm$ and $p6mm$, respectively).
This deflection of the trajectories towards the diagonal is visibly smaller for the case of hexagonal cylinders, where chains are deflected by a smaller angle due to the higher coordination (or symmetry).

For both square and hexagonal phases, the cylindrical domain outlined by the IMDS becomes increasingly warped away from a circular shape with either increasing with $f$ or $\epsilon$ (Fig.~\ref{fig:traj_wedges}C, G).  Increasing both parameters is expected to increase the relative importance of the stretching free energy of matrix (B) blocks relative to the core (A) blocks and the IMDS surface energy, and have been argued to lead to ``quasi-faceted'' IMDS shapes akin to the straight-path SST assumptions ~\cite{Grason2004, Grason2006}.  For a given set of parameters, this IMDS faceting is more obvious for the square packing, presumably due to the larger variation of IMDS-to-outer terminal distance traversed by which is mitigated by deforming the IMDS such that the matrix domain approaches a more uniform thickness.  

Underlying the more obvious changes of IMDS shape with increasing $\epsilon$ or $f$ are more subtle changes in trajectories.  We first analyze the kink angles $\beta$ around the IMDS in Fig.~\ref{fig:bend_horizontal}.  Fig.~\ref{fig:bend_horizontal}A and E compare the kinking as function of angular position at the IMDS, which reverses sign across the mirror planes separating asymmetric wedges (at $\theta_{\rm max} = 45 ^\circ$ and $30 ^\circ$ respectively).  Interestingly, the sign of the kink angle $\beta$ within a fundamental domain ($0 \leq \theta < \theta_{\rm max}$) depends on $\epsilon$, with relatively small negative values at low $\epsilon$ and large positive values at high $\epsilon$. As such, both morphologies exhibit an inversion in the sign ({\it or direction}) of kinking, during which the kink angle vanishes, on average (at $\epsilon \simeq 1.45$ and $\epsilon \simeq 1.75$ for $p4mm$ and $p6mm$, respectively).  Curiously, both morphologies exhibits kink angles that alternate in sign {\it within a fundamental domain} for intermediate values of $\epsilon$ around the inversion point, as shown for $p4mm$ in Fig.~\ref{fig:bend_horizontal}A.

Importantly, the general variation of mean kink angle on $\epsilon$ is robust, confirmed for broad ranges of $\chi N$ and $f$ (see Fig.~\ref{fig:bend_horizontal}C,D and G,H).  Perhaps intuitively, we observe a much large degree of kink in square over hexagonal cylinders, accounting for the better close packing geometry of cylinders in hexagonal over square lattices.

The negative kink angles at small $\epsilon$ are seemingly consistent with the kinked-path ansatz~\cite{Olmsted1998}, where stiffer A blocks prefer a uniformly radial packing, and deflect towards an interstitial corner (next nearest neighbor direction) at the IMDS in order to fill the B domain at uniform density.  Surprisingly, this degree of negative tilt is relatively small compared to a much more prominent positive kink at large $\epsilon$.  This effect, which might be called a ``counter-kinking" arrangement, can be rationalized heuristically by the tendency of the stiffer block to intersect the IMDS in a nearly-normal orientation in order to minimize the cost of stretching (i.e. associating to the shortest possible path to the IMDS in stiffer domain).  There is a smaller penalty for the less-stiff block to meet the IMDS off-normal, along a longer distant path that presumably has to absorb the cost of packing frustration, and results in a kinked trajectory that becomes increasingly kinked for more extreme values of conformational asymmetry $\epsilon$.
Since the cylinder and matrix domains invert in stiffness as $\epsilon$ is tuned through 1, this argument predicts an inversion of the kink angle; a similar tendency has been seen in the triply-periodic network phases~\cite{Dimitriyev2023}.  Relative to low $\epsilon$, the degree of positive counter-kink for large $\epsilon$ is enhanced due to simultaneous faceting of the IMDS, which becomes more parallel to the outer terminal boundary (i.e. the square and hexagonal Voronoi cells), requiring even greater deflection to the nominally radial inner block trajectories (see Fig.~\ref{fig:traj_wedges}D, H).

\begin{figure*}[t]
\centering
\includegraphics[width=\textwidth]{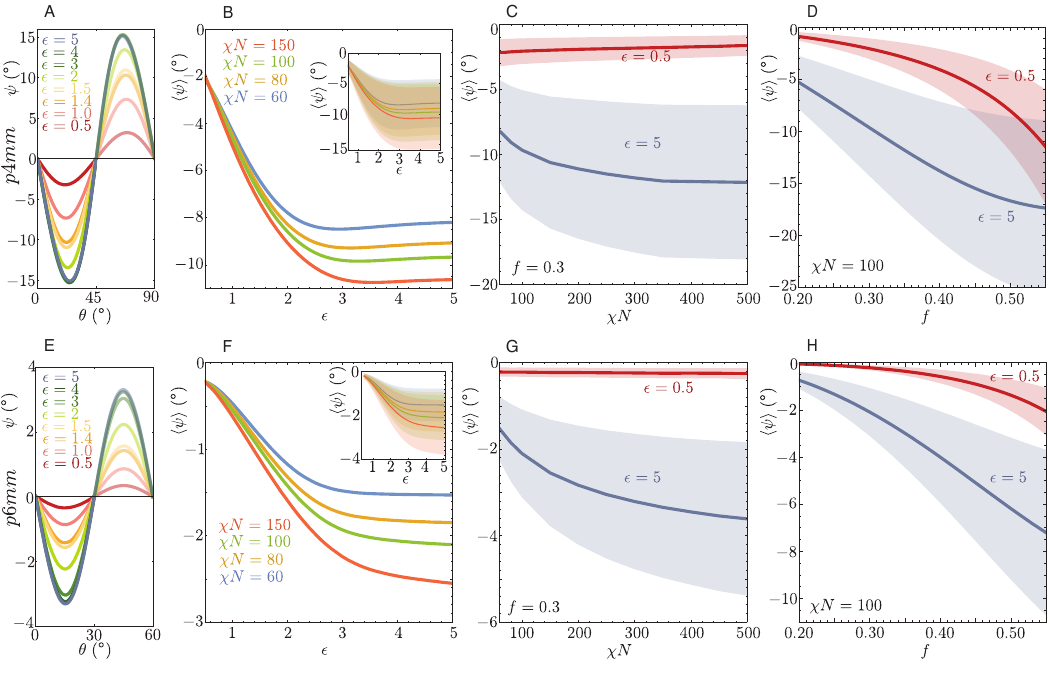}
\caption{\label{fig:tilt_horizontal}(A) Tilt angle as a function of polar angle $0 \leq \theta \leq 90 ^\circ$ for the square cylinder phase at $\epsilon = 0.5 - 5.0$ (fixed $\chi N = 100$, $f=0.3$). (B) Average of the tilt angle, $\langle \psi \rangle$, taken over a fundamental domain as a function of $\epsilon$ for $\chi N = 60-150$ (fixed $f=0.3$) showing that the tilt angle is negative for all $\epsilon$. Inset shows the variation in the tilt angle, where the shaded region extends for a single standard deviation on either side of the mean. Finally, tilting angle statistics for extreme values of conformational asymmetry ($\epsilon = 0.5$ and $5.0$) are shown as a function of (C) $\chi N$ and (D) $f$, confirming the consistency of the sign of the tilt angle. (E)-(H) show these data for the hexagonal cylinder phase ($p6mm$).}
\end{figure*}

This scenario is consistent with the trends observed from tilt angle $\psi$ of mean trajectories with respect to the IMDS, shown in Fig.~\ref{fig:tilt_horizontal}.  Here, we observe, for both symmetries and all conditions, that the sign of $\psi$ is always the same (negative), and the inclination increases with {\it both} $\epsilon$ and $f$.  This trend is consistent with the increasing IMDS faceting, which varies from relatively round (low $ \epsilon $, $f$) to relatively polyhedral (high $ \epsilon $, $f$), driven the by prerogatives of the stiffer block.  Hence, even while the chain trajectories remain within a few degrees of radial, deformation of the IMDS away from circular implies they meet it at tilt with increasing quasi-polyhedral warping, as shown in Fig.~\ref{fig:traj_wedges}C,D and G,H. 

Taken together, comparison of the similar ranges of kink and tilt angles in Figs.~\ref{fig:bend_horizontal} and ~\ref{fig:tilt_horizontal} suggest that to a large extent, the trajectory at the IMDS is determined by the warping of its shape from circular towards polyhedral, combined with the effect of elastic asymmetry to orient the stiffer block trajectories normal to the IMDS.  Additionally, the degree of kinking, while measurable, remains fairly small for the most stable columnar state of hexagonal cylinders.  For both morphologies, we find evidence that the degree of tilt and kink tends towards asymptotic saturation in the $\chi N \to \infty$ limit, suggesting the features and trends analyzed at finite segregation bear hallmarks of a well-defined SST packing limit.

% \begin{figure*}[h!]
% \centering
% \includegraphics[width=7in]{figures/bend.png}
% \caption{\label{fig:bend}: Bend. }
% \end{figure*}

% \begin{figure*}[h!]
% \centering
% \includegraphics[width=7in]{figures/tilt.png}
% \caption{\label{fig:tilt}: Tilt. }
% \end{figure*}

\subsection{\label{sec:2D_p4p6} Trajectories near the their termini: Packing in Stretched ($c2mm$) Cylinders }

Next, we turn to the shape of domain edges, i.e.~the {\it terminal boundaries}, of columnar phases.  For BCP and other amphiphile assemblies, the ends of core (A) and matrix (B) domains of columnar phases are often assumed take simplified forms.  The contact surface between opposing brushes in the outer boundary is most often approximated by the 2D Voronoi cell of the packing, while inner chain trajectories are assumed to extend to a central line extending along the centroid of the 2D cross-section of that cell.  As discussed previously~\cite{Reddy2021, Reddy2022}, these seemingly intuitive assumptions are likely to fail to describe the domain geometry in many conditions, specifically because they do not reflect how changes in IMDS shapes impact the association of BCP chains the IMDSs.  That is, Voronoi cells describe only the set of points closest to a given generating point (usually a Wyckoff position of 2D crystal), not necessarily the points where chains are most likely to associate to one IMDS or another.  Previously it was argued that the {\it medial map}~\cite{Reddy2021}, that maps all points in the morphology onto the closest point on the IMDSs~\cite{Schroder2003, Schroder-Turk2006}, would therefore provide a better (albeit purely geometric) proxy to an association map for a given IMDS shape.  

Here we analyze the association maps and their terminal boundaries directly from SCF solutions for a set of columnar morphologies with non-trivial symmetry.  In particular, we compute a family of solutions for fixed parameters --- $f =0.3$, $\epsilon = 1$, and $\chi N = 100$ --- for the $c2mm$ space group.  These solutions correspond to centered rectangular unit cells, including two columns per cell, in general parameterized by unit cell parameters $\ell_x$ and $\ell_y$.  We define the aspect ratio $\lambda \equiv \sqrt{3} \ell_x/\ell_y$ in terms of the distortion away from $p6mm$-like packing at $\ell_x/\ell_y=1/\sqrt{3}$, which has higher, sixfold symmetry.

\begin{figure*}[t]
\centering
\includegraphics[width=\textwidth]{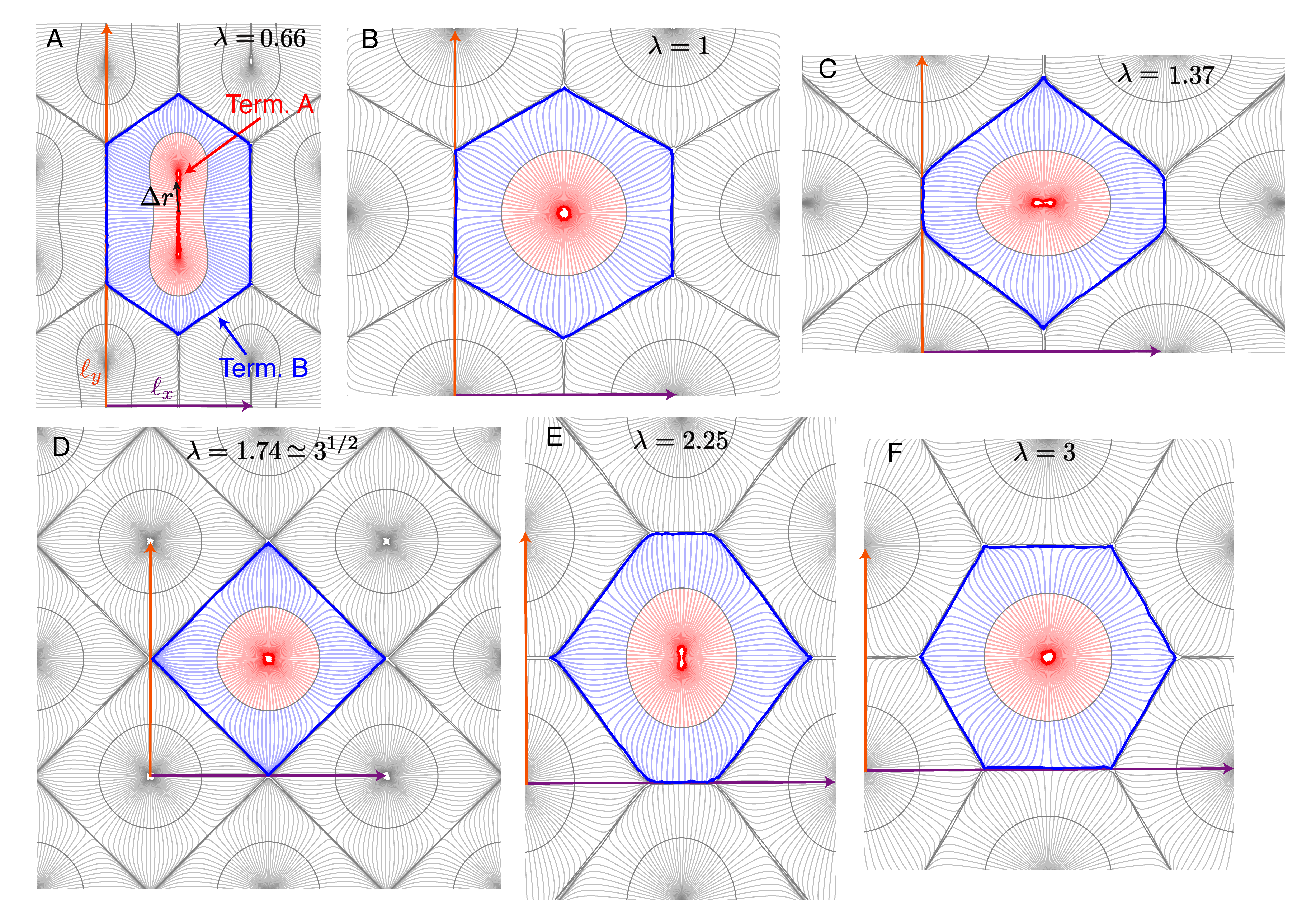}
\caption{\label{fig:c2mm_prog} Domain morphologies, chain trajectories, outer terminal boundaries (blue) and inner terminal boundaries (red) for stretched cylinders along the deformation path (steps A - F). The hexagonal cylinder structure in step B is transformed to an equivalent hexagonal cylinder structure at step F (rotated by $90 ^\circ$) by passing through a square phase at step D. Purple and orange arrows delineate the $x$- and $y$-edges, ($\ell_x$, $\ell_y$) respectively, of the unit cell at each step.}
\end{figure*}

For each $\lambda$ we consider SCF solutions with minimal free energy cross-sectional area but a fixed aspect ratio.  We consider the range $0.66 \leq \lambda \leq 3$, spanning a range where the nearest neighbor direction in hexagonal packing is relatively compressed ($\lambda<1$) or stretched ($\lambda>1$).  For small distortions from $\lambda =1$, the cross-sectional IMDS shapes become evidently eccentric and approximately elliptical with major axes along the stretch direction.  Fig.~\ref{fig:c2mm_prog} shows a sequence SCF solutions with corresponding association maps and terminal boundaries.  Notably, for $\lambda < 1$ the domain cross-section becomes increasingly stretched, far beyond elliptical, for aspect ratios much lower than 1 (Fig.~\ref{fig:c2mm_prog}A).  In contrast, increasing $\lambda>1$ leads to a more complex evolution of domain anisotropy.  This is because $\lambda = \sqrt{3}$ (Fig.~\ref{fig:c2mm_prog}D) corresponds to equal unit cell dimensions, commensurate with a square ($p4mm$) packing.  And stretching even further past the square packing to $\lambda = 3$, the structure again returns packing to hexagonal ($p6mm$) but with neighbor directions rotated by $30 ^\circ$ relative to $\lambda =1$ (Fig.~\ref{fig:c2mm_prog}F). Hence, at these special aspect ratios, the IMDS shapes are fairly circular (or at least, consistent with 6- or 4-fold symmetry) while at intermediate points (Fig.~\ref{fig:c2mm_prog}C, E) the domains exhibit evident anisotropic shapes.

The effect of domain anisotropy has an obvious effect on the underlying chain packing, most evident in the geometry of the inner (A) terminal boundary.  Crudely speaking, the terminal boundaries of {\it isotropic domains} appear point-like, to our ability to resolve them via discrete meshes (see Appendix~\ref{app:term}), while for anisotropic shapes, these boundaries {\it spread} along the longer axis of the domain shape.  The spreading of the terminal boundary reflects the fact that the chain trajectories do not focus to a central point in the cross-section, and to some extent pack in quasi-lamellar fashion at the core of the domain.

\begin{figure}[h!]
\centering
\includegraphics[width=3in]{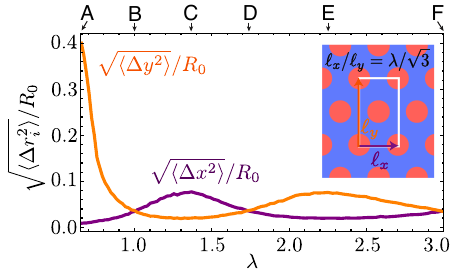}
\caption{\label{fig:c2mm_measure} Structure of the family of continuously stretched cylinders, parameterized by stretch factor $\lambda$. Inner terminal boundary shape, quantified by the r.m.s.~spread of points around the domain center along the x-direction ($\sqrt{\langle \Delta x^2\rangle}$, purple) and the y-direction ($\sqrt{\langle \Delta y^2\rangle}$, orange), in units of the effective IMDS radius $R_0$. Inset relates the aspect ratio $\ell_x/\ell_y$ of the $c2mm$ unit cell to $\lambda$. 
}
\end{figure}

The extent of this line-like inner terminal boundary provides a measure of quasi-lamellar packing in the anisotropic columnar phase.  To characterize the shape of the inner (A) terminal boundary, we compute its r.m.s.~deviation from the cell centroid in both directions, $\langle (\Delta x)^2\rangle^{1/2}$ and $\langle (\Delta y)^2\rangle^{1/2}$, and normalize those by $R_0$, defined as the radius of circular equal area to the A-subdomain (i.e.~$\pi R_0^2 = f/2 \ell_x \ell_y$).  Results for the the r.m.s.~extent of the A terminal boundary in both directions are plotted in Fig.~\ref{fig:c2mm_measure} as a function of $\lambda$.  These show that for the cases of hexagonal ($\lambda=1$, 3) or square ($\lambda=\sqrt{3}$), $x$ and $y$ dimensions of terminal boundary are equal (and of small magnitude), consistent with point-like shape, although finite mesh limits the ability to resolve any potentially finer scale features.  Away from the these points, these alternate dimensions become unequal, by amounts that are consistent with the visibly anisotropic IMDS shapes.  Note, that while the terminal boundary is likely line-like for most of these anisotropic shapes, the minimal value of the scaled r.m.s.~dimensions of the terminal boundary is never lower than $\sim 10^{-2}$, reflecting the inherent limits to imposed of finite grid resolution (as discussed in Appendix~\ref{app:term}).  The observed changes in these dimensions are at least an order of magnitude larger than this scale, indicating that at least for sufficiently anisotropic of domains, analysis of the singularities of association map reconstructed from SCF is nevertheless able to illuminate these otherwise ``hidden'' features of chain packing and their dependence on anisotropic domain shapes.

\section{\label{sec:3D}Three-dimensional morphologies: Network and micellar crystals}

We now turn to 3D complex morphologies, which are often considered to be subject to even larger degrees of packing frustration than their 2D columnar counterparts~\cite{ACShi2021}.  Having shown that the anisotropy of 2D domains has a prominent impact on underlying chain trajectories, most notably the spreading of the terminal boundaries, here we focus on the terminal boundaries of two classes of 3D morphologies:  bicontinuous networks and 3D crystalline arrangements of quasi-spherical (micelle-like) domains. In both cases, the nature and shapes of the terminal boundaries and their role in describing the thermodynamics of packing frustration has been the subject of considerable debate and speculation for decades.  Here, we analyze the terminal boundaries directly from the SCF description of BCP conformations, and compare it to the previously invoked proxies for the shape of domain edges, including Voronoi and medial geometry.

\subsection{\label{sec:DG}Bicontinuous double-gyroid networks}

Terminal geometry plays a crucial role in stabilizing triply-periodic network phases.
In particular, the stability of the double gyroid (DG) with respect to lamellae and columnar morphologies has long been thought to be due to its ability to relieve ``packing frustration''~\cite{Matsen1996}.
However, the prevailing picture relied on a simplification of terminal geometry wherein the inner terminal boundary consists of the 1D skeletal graph, with the outer terminal boundary resembling the gyroid minimal surface~\cite{Prasad2018}.
The skeletal graph approximation for the inner terminal boundary proved problematic for investigations of packing frustration based on strong-segregation theory (SST), as it places extreme constraints on A-block chain packing~\cite{Olmsted1994, Olmsted1995, Olmsted1998, Likhtman1994, Likhtman1997}.
Recently~\cite{Reddy2022, Dimitriyev2023}, it was shown via a ``medial surface'' construction of SST (so-called medial-SST) that the inner terminal boundaries are better approximated by twisting ribbon-like surfaces that contain the skeletal graph, yet relax constraints on chain packing without violating the space-filling constraints required of a polymer melt.  Crucially, it was shown in the SST limit, that entropic relaxation due to ``spreading'' of terminal ends of the tubular block (A-block) away from the confines of the 1D skeleton onto a web-like terminal surface is necessary to account for the equilibrium stability of DG intermediate to (hexagonal) columnar and lamellar morphologies for diblock melts.  Here, we test this  {\it medial ansatz} for packing in DG morphologies by direct analysis of the terminal boundaries from SCF solutions.

\begin{figure}[h!]
\centering
\includegraphics[width=3in]{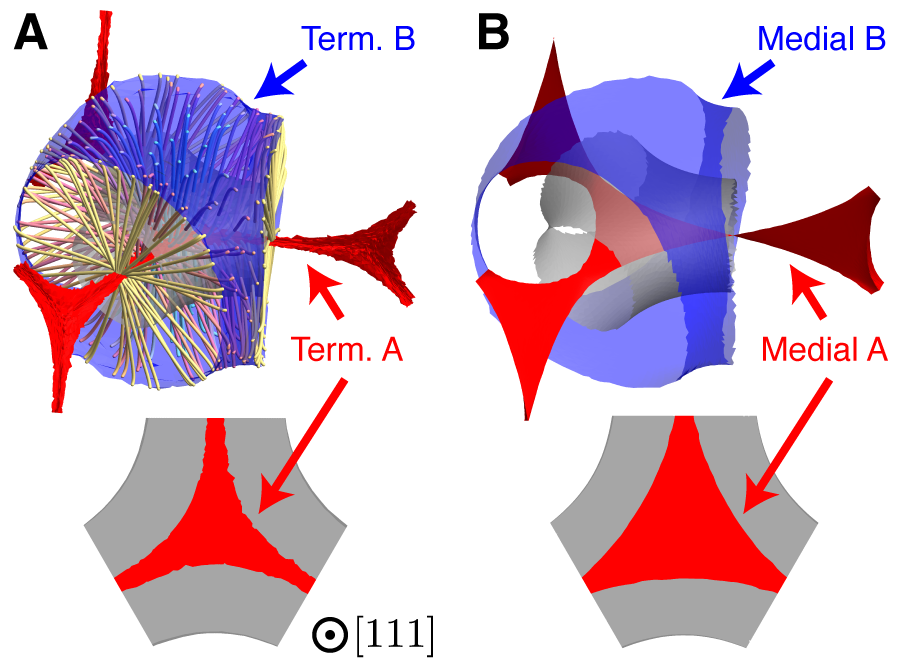}
\caption{\label{fig:gyroid} ``Mesoatom'' unit of the double gyroid network phase with inner and outer terminal boundaries shown in red and blue; the IMDS is shown in gray. There is additional inner terminal ``web'' shown outside of the mesoatom that twists between neighboring nodes. (A) shows SCF-computed ($f = 0.29$ and $\chi N = 50$) terminal boundaries and a selection of trajectories. (B) shows the medial set model of terminal boundaries. Comparisons of the inner terminal boundary geometry along the $[111]$-direction are shown below.}
\end{figure}

Using the terminal map $\bm{\alpha}$ extracted from SCF trajectories, we find a set of terminal boundaries for a DG morphology with $f = 0.29$ and $\chi N = 50$, shown in Fig.~\ref{fig:gyroid}A.
Here, we focus on a fundamental unit of the double gyroid centered about the network nodal regions (i.e.~the 16b Wyckoff positions) as an analog of the Voronoi cells in columnar and spherical phases.
As they are bounded by a terminal boundary that wraps a single domain, these have been dubbed the ``mesoatoms'' of the DG assembly, and are taken to represent a basic unit of self-assembly~\cite{GrasonThomas2023}.
The inner terminal boundary consists of a nearly-flat surface with trihedral coordination that twists by $70.5^\circ$ about edges connecting neighboring nodes, similar to the inner medial web shown for comparison in Fig.~\ref{fig:gyroid}B.
The extended flat sections of the inner terminal web indicate regions of quasi-lamellar chain packing, with chain trajectories in the vicinity oriented roughly parallel to each other. About the thin strut region and along the boundary of the inner terminal web, trajectories extend in a radial pattern, indicating quasi-cylindrical chain packing. This hybrid of lamellar and cylindrical packing is thought to rationalize the stability of the double gyroid phase intermediate to lamellar and columnar phases, which has been supported by the medial model of terminal boundaries for the double gyroid and other network phases~\cite{Dimitriyev2023}.
As demonstrated by the projection of the terminal webs along the $[111]$-direction in the bottom panels of Fig.~\ref{fig:gyroid}, the SCF-computed inner terminal boundary is similar in gross shape to medial surface generated by the same IMDS, but is slightly reduced in dimensions relative the medial set, where the projected area of the surface shown in the inset of Fig.~\ref{fig:gyroid}A is roughly 50\% less than that of the corresponding medial surface in the inset of panel B.
We attribute the reduction in size of the SCF inner terminal boundary compared with the inner medial surface to the fact that trajectories are in general not normal to the IMDS, and exhibit at least a modest degree of tilt and kinking (see ~\cite{Dimitriyev2023}), which is also evident in Fig.~\ref{fig:gyroid}A.

\begin{figure}[h!]
\centering
\includegraphics[width=3in]{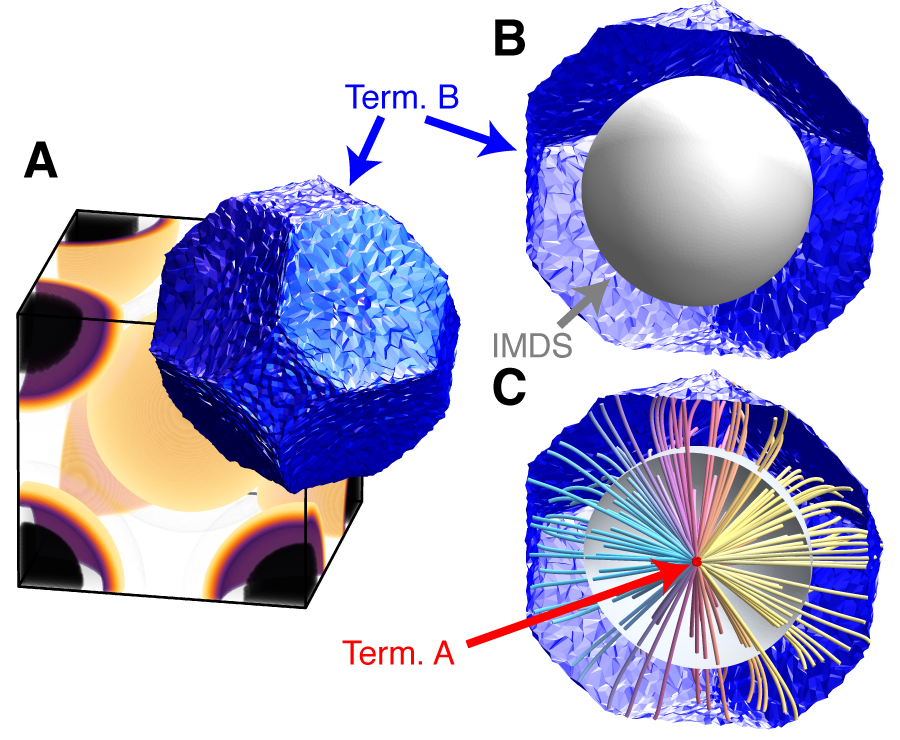}
\caption{\label{fig:bcc} (A) Plot of A-block density field $\phi_{\rm A}(\mathbf{r})$ in a unit cell of the BCC sphere phase with single mesoatomic cell bounded by an outer terminal boundary (blue), with $f = 0.29$ and $\chi N = 50$. (B) Cutaway of the outer terminal boundary shows an IMDS (gray). (C) Cross-section of the IMDS shows the point-like inner terminal boundary (red), along with a selection of trajectories.}
\end{figure}

\subsection{\label{sec:spheres} Classical and Frank-Kasper sphere phases}

In contrast to the network morphologies, sphere morphologies share a compact domain structure that are a 3D analog of the 2D cross-sections of the columnar morphologies already discussed. These sphere phases are effectively crystalline packings of micelle-like domains, warped in shape by the lower-symmetry constraints of their inter-domain arrangement~\cite{Grason2006,Reddy2018, ACShi2021}.  Here, we analyze SCF solutions at $\chi N = 50$ and $f=0.29$, which is above the core composition window for sphere phases for elastically-symmetric chains, but within the window where stiff matrices stabilize them for high $\epsilon$, notably the Frank-Kasper phases~\cite{Dorfman2021}. 

The most generically stable of the simple sphere phases, the BCC packing, is shown in Fig.~\ref{fig:bcc}, with rendered chain trajectories and terminal boundaries. Owing to the single mesoatom of the BCC crystal along with symmetry constraints, the outer terminal boundary, which we refer to as its {\it terminal cell}, appears to conform closely to its Voronoi cells, which are truncated octahedra (note, we do not attempt to resolve the possible curvature of the terminal cells).  As shown in Fig.~\ref{fig:bcc}B, the IMDS enclosed by the outer terminal boundary is nonetheless highly spherical (here, shown for $f = 0.29$ and $\epsilon = 1.0$ at $\chi N = 50$).
The association of points between the faceted outer terminal boundary and the round IMDS requires curved chain trajectories, as shown in Fig.~\ref{fig:bcc}C, terminating in a small, roughly point-like inner terminal boundary, consistent with the conventional assumption of radial packing.

\begin{figure*}[t]
\centering
\includegraphics[width=\textwidth]{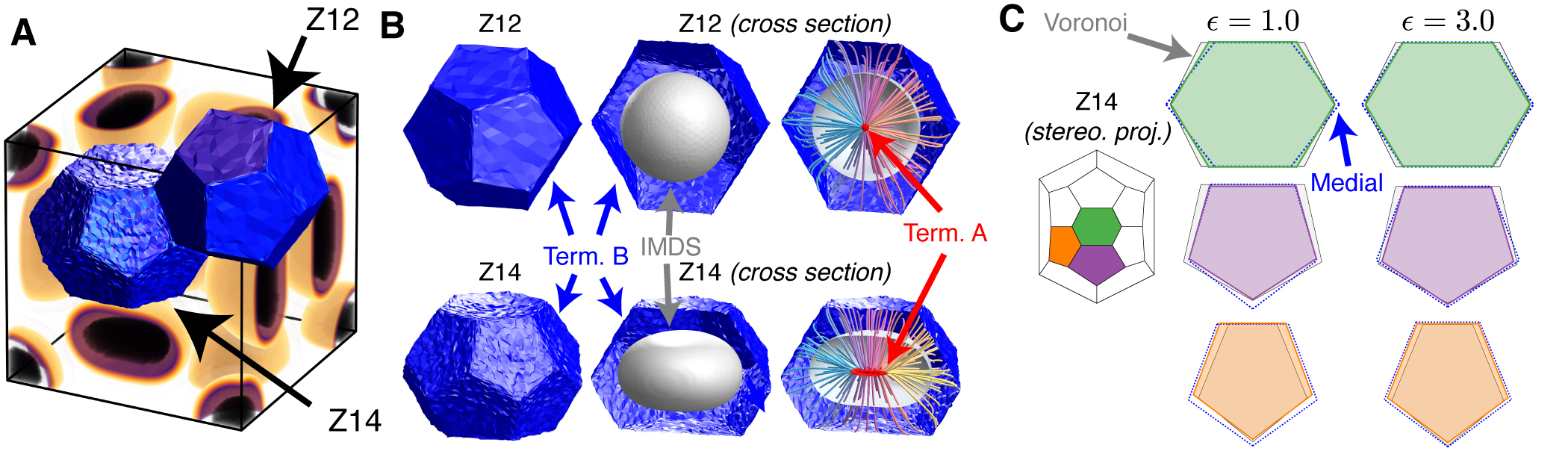}
\caption{\label{fig:a15_cells} (A) Plot of A-block density field $\phi_{\rm A}(\mathbf{r})$ in a unit cell of the A15 sphere phase with Z12 and Z14 mesoatomic cells bounded by outer terminal bounadries (blue), with $f = 0.29$ and $\chi N = 50$. (B) Individual Z12 cells (top) and Z14 cells (bottom) with cutaways showing the IMDSs (gray) and inner terminal boundaries (red), along with a selection of trajectories. (C) Planar projections of polygonal facets from different models of outer terminal boundaries, with Voronoi cell polygons shown in gray and medial boundary polygons shown as blue, dashed lines. Faces correspond to those labeled on the stereographic projection of the Z14 cell, with the hexagonal Z14-Z14(6) boundary in green, the pentagonal Z14-Z12(5) boundary shown in purple, and the pentagonal Z14-Z14(5) boundary shown in orange. Comparisons between $\epsilon = 1$ and 3 are shown.}
\end{figure*}

The complex Frank-Kasper sphere phases require multiple, distinct mesoatom units that exhibit different symmetries according the symmetry-inequivalent Wyckoff positions~\cite{Dorfman2021}.
Of these, the A15 (shown in Fig.~\ref{fig:a15_cells}A-B extracted from chain trajectories for $\epsilon=1$) is arguably the simplest example with only two inequivalent mesoatomic units.  It also plays a particularly interesting role, as the cell boundaries resemble the Weaire–Phelan foam~\cite{Weaire1994}, which has a smaller area per volume than the BCC lattice, leading to windows of stability despite its relatively complex structure~\cite{Grason2003, Grason2004, Xie2014, Bates2019}.
It consists of two mesoatoms, labeled by coordination: the Z12, centered at the 2a Wyckoff sites and the Z14, centered at the 6c Wyckoff sites of the $Pm\bar{3}n$ space group.
These coordination numbers also describe the number of faces in the terminal polyhedra that form the boundaries of the cells housing each of the mesoatoms.
As shown in Fig.~\ref{fig:a15_cells}B, the Z12 cells have dodecahedral outer terminal boundaries, consisting of identical pentagonal facets, whereas the Z14 cells have boundaries consisting of a pair of hexagonal facets and twelve pentagonal facets.  It is generally observed for Frank-Kasper phases of soft matter assemblies, that the shapes of the mesoatomic units can be quite distinct.
For A15, Z12 cells enclose nearly-spherical IMDSs, while the Z14 cells enclose fairly oblate quasi-ellipsoidal IMDS shapes, visibly ``squashed'' along the stacking directions for neighbor Z12 cells~\cite{Reddy2018, Reddy2021, Collanton2022}.  This different symmetry is reflected in the underlying packing in the core, where Z12 is generically observed to maintain a point-like inner terminal boundary, while the discoidal Z14 domain exhibits a disk-like terminal surface that spreads along the wider dimension of the domain.  
Much like the quasi-lamellar packing seen in the double-gyroid, a bundle of nearly-parallel trajectories emerge from the Z14's terminal disk, resulting in similar quasi-lamellar packing along the stacking direction, with radial packing along the edges.
This difference in packing motifs within the same structure leads to asymmetry in the shape of the outer terminal boundary, namely the polygonal facets that separate neighboring cells.

In Fig.~\ref{fig:a15_cells}C we compare the detailed shapes of outer terminal cells to two other models of boundaries between quasi-spherical domains:  medial and Voronoi cells~\cite{Reddy2021}. In particular, we directly compare the shapes of the distinct polygonal faces (all of which belong to the Z14 cell), for two values of conformatial asymmetry, $\epsilon=1$ and 3.  The hexagonal faces (here referred to as Z14-Z14(6), shown in green) separate neighboring Z14s along the stacking direction; a collection of four pentagonal faces (Z14-Z12(5), purple) separate Z14 and Z12 cells; the remaining eight pentagonal faces (Z14-Z14(5), orange) separate Z14 cells off of the stacking direction.
Due to the slight curvature in the terminal faces, for a given face, we determine an approximate tangent plane (by averaging over all of the normal vectors of that face), and then project the face onto that tangent plane; we then fit the average shape of the face to a polygonal boundary.  Through this procedure, we find that the SCF-computed terminal cell, while resembling Voronoi cells, differ in size and shape, with smaller Z14-Z14(6) and Z14-Z12(5) faces and larger Z14-Z14(5) faces.
Compared to the Voronoi cell, the terminal cell is closer in shape to the outer medial surface generated by the IMDS, which is consistent with the notion that the medial map is a close approximation of the association map as it minimizes the cost of chain stretching~\cite{Reddy2021}.  Interestingly, as the matrix phase is made stiffer ($\epsilon = 3$ shown), the SCF-computed terminal cell approaches the shape and size of the outer medial surface even more closely.
This suggests that prerogatives of the stiffer matrix block restructure the chain packing in a way that brings it somewhat closer ``medial packing,'' an effect which has been predicted for DG networks~\cite{Reddy2022, Dimitriyev2023}.

\begin{figure*}[t]
\centering
\includegraphics[width=\textwidth]{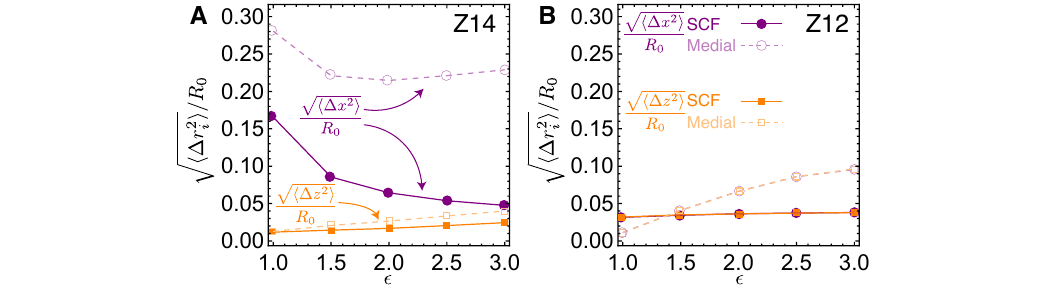}
\caption{\label{fig:a15_termA} Comparisons of A15 inner terminal boundary geometry computed with SCF (solid lines and symbols) against medial set approximations (dashed lines, open symbols) for (A) Z14 cells and (B) Z12 cells as a function of $\epsilon$. Variance in the distribution of points about the nodal center, $\langle \Delta r^2_i \rangle$ are plotted in reference to the effective IMDS radius $R_0$, with the $x$-component of the variation, $\sqrt{\langle \Delta x^2 \rangle}/R_0$, shown in purple and the $z$-component of the variation, $\sqrt{\langle \Delta z^2 \rangle}/R_0$, shown in orange. Here, $R_0$ is the radius of a sphere with equal volume to the (A-block) domain core.}
\end{figure*}

The variations in outer terminal boundary shape with changing conformational asymmetry $\epsilon$ accompany changes in the geometry of the inner terminal boundary.
As shown in Fig.~\ref{fig:a15_termA}A, the discoidal inner terminal boundary of the Z14 presents as a significant anisotropy in its shape, which is narrow (surface-like) along the local stacking direction (here, the $z$-axis) and spread out along the transverse directions (here, the $x$- and $y$-axes).
As $\epsilon$ increases, the spread of points along the disk narrows, resulting in a decrease in $\sqrt{\langle \Delta x^2 \rangle}$ (equivalent to $\sqrt{\langle \Delta y^2 \rangle}$, which is not shown), while the disk thickens, indicated by an increase in $\sqrt{\langle \Delta z^2 \rangle}$.
This marks a trend wherein the terminal disk becomes more isotropic and point-like, approaching the dimensions of the Z12 inner terminal boundary, shown in Fig.~\ref{fig:a15_termA}B, which has a radius of gyration that is $\sim 5-7\%$ of the effective radius of the IMDS.
By comparison, the inner medial surfaces for the Z14 and Z12 cells, while similar in size and showing similar trends near $\epsilon \simeq 1$, attain significantly different dimensions for larger values of $\epsilon$.

Notably, this transition from quasi-discoidal to radial chain packing from low- to high-$\epsilon$ in the squashed FK domains was previous suggested based on observations of IMDS shape variation with $\epsilon$~\cite{Reddy2018, Dorfman2021}, but here we resolve it directly from the changes in trajectories encoded the statistical descriptions of BCP conformations available in SCF.  As the matrix block becomes stiffer and the morphology is increasingly dominated by the need to minimize the stretching cost of the matrix, the stretching cost of chains in the spherical core has a smaller relative impact on the total free energy.
Thus, for large enough $\epsilon$ the cost of maintaining an anisotropic cell outweighs the benefit of forming quasi-lamellar regions and a discoidal inner terminal boundary of Z14.

We anticipate that similar transitions in sub-domain chain packing underlie the structure of a broader family of Frank Kasper phases, as well as their related dodecagonal quasicrystalline cousins.  We expect further, that differences between medial and terminal cell shapes are likely to be even more pronounced in phases like C14 and C15, where the relative volumes between distinct mesoatoms is much greater than the case for A15, here found to exhibit $\lesssim 20 \%$ difference between Z14 and Z12.

\section{\label{sec:conclusion}Discussion and concluding remarks}

% BIG SUMMARY/TAKEAWAYS
% extracted average chain trajectories from SCF calculations... "hidden information"
% found that chains generally tilt/kink to fill space
% mapped out domain structure, notably locations and geometry of terminal boundaries
% comparisons between extracted terminal boundaries and medial geometry

% WHAT HAVEN'T WE STUDIED...
% more details of networks phases
% tilt/kink relationships + dithering zone; (focus on regions where phases are stable)

In this study, we employed the statistical description of fluctuations of BCP chains via SCF to extract average chain trajectories and analyze detailed features of the subdomain packing, features which are often considered be ``invisible'' in this formalism.  From analysis of the chain trajectories, we found that chain tilt and kinking are generic features of packing within curved morphologies.
These behaviors bear some characteristic behavior of previously-untested ansatz and also show some surprising new trends.
We additionally extracted terminal boundaries from these SCF-computed chain trajectories, features of block copolymer domains that have been recognized as playing a key role in the formation of complex and frustrated morphologies, but again, have been otherwise ``invisible'' to direct study.  
Typically modeled using oversimplified proxies for domain boundaries, (e.g.~Voronoi tessellations, skeletal graphs, and more recently, medial surfaces), here we demonstrate how these can be rigorously extracted directly from the statistical description of chain degrees of freedom in SCF.   Finally, for two examples of 3D morphologies, we tested and broadly confirmed some recent conjectures about the connections between chain packing and the medial geometry of domains~\cite{Reddy2021, Reddy2022, Dimitriyev2023}.  

We note that this analysis for packing features of finite-$\chi N$ raises a number of open questions, even for the restricted cased on AB diblock BCP melts, for example whether terminal boundaries of other complex morphologies, like the double-diamond and double-primitive networks, are indeed composed of multiple ``leaves'' joining at finite angles, as suggested by corresponding medial surface~\cite{Dimitriyev2023}.  Moreover, a more mechanistic understanding of the apparently complex interplay between distinct, multiple ``modes'' of responding to packing frustration (e.g.~chain tilt, kink and combined shapes of IMDS {\it and} terminal boundaries) and how these vary with BCP parameters remains to be explored.  

% EXTENSIONS
% multiple blocks, different architectures
% blends (what happens to homopolymers... where do they go?)
% what about different branches of a complex molecules, eg dendrimer
% brushes... interpenetration
% connections to simulations w/ multiple chains (md/mc)? Poornima's paper?
% experiments? how to see? perspective... nature paper...

Access to average chain trajectories within SCF gives a direct microscopic view of chain packing that has only previously been accessible to molecular simulation methods, such as molecular dynamics and Monte Carlo simulations~\cite{Martinez2006, Martinez-Veracoechea2009, Prasad2017, Shen2018}, while ensuring that equilibrium conditions are maintained.
While such simulation techniques can reveal rich subdomain structures~\cite{Padmanabhan2016}, general and robust approaches to quantify spatially resolved mean-trajectories and terminal boundaries from ensembles of fluctuating chain configurations remains an open challenge.
Related to this, the inclusion of homopolymer has important consequences for the thermodynamics of a BCP phase and the distribution of homopolymer has been widely studied, in particular for its purported effects to relieve the costs of packing frustration~\cite{Matsen1995, Matsen1996, Martinez-Veracoechea2009, Cheong2020, Xie2021}.  Notably, the packing of homopolymer chains amongst BCPs can be addressed via suitable extension of the approach introduced above for diblock melts.  
Given resolution limits of microscopy, resolving information about chain trajectories and terminal boundaries remains an outstanding experimental problem, yet advancements in sophisticated 3D microscopy (namely ``Slice-and-View'' SEM) and reconstruction techniques that have already shed light on subdomain geometry~\cite{Feng2019,Reddy2021} and refinements of such methods may make this feasible.
Beyond linear diblocks, the approach presented in this manuscript may be readily applied to more complex block copolymer architectures, such as branched polymers and even bottlebrush polymers and dendrimeric copolymers~\cite{Matsen2012, Polymeropoulos2017}.
These complex architectures involve block-specific orientational order parameter fields, leading to rich multiplexed trajectory information, which may give rise to exotic packing structures, such as nested terminal boundaries.
Finally, trajectory information obtained from SCF calculations of polymer brushes can help address outstanding questions about brush organization and intrabrush segregation~\cite{Minko2003} and even interpenetration between brush-coated nanoparticles~\cite{Midya2020}.

\section*{Acknowledgements}

The authors gratefully acknowledge valuable discussions with A.~Reddy and E.~Thomas.  This research was supported by the U.S. Department of Energy (DOE), Office of Basic Energy Sciences, Division of Materials Sciences and Engineering, under award DE-SC0022229. SCF computations and chain trajectory analysis were carried out on the Unity Cluster at the Massachusetts Green High Performance Computing Center.

\section*{Conflicts of interest}
The authors declare no conflicts of interest.

\section*{Data availability}

The analysis code that was used for this study is openly available in UMass Amherst ScholarWorks at \url{https://doi.org/10.7275/1b2p-q547}.

\appendix

\section{\label{app:op} Order parameter calculations}

Calculation of the polar order parameter field $\mathbf{p}_{\alpha}(\mathbf{r})$ requires prior calculation of the chain end distribution functions $q_{\pm} ({\bf r}, n)$, which can be exported from the open-source PSCF software used for our SCF calculations~\cite{Arora2016} (available at \url{https://pscf.cems.umn.edu/}).
The export feature writes a pair of files (one for $q_-$ and the other for $q_+$) for each of the two blocks considered here; in general, for multi-block architectures, a separate pair of files is exported for each block.
We have written a Python script (available at \url{https://doi.org/10.7275/1b2p-q547}) that performs the finite difference and numerical integration necessary to the calculation of $\mathbf{p}_{\alpha}(\mathbf{r})$.
To perform the calculation for a given block $\alpha$, the user supplies the pair of chain end distribution files, along with simulation parameters such as unit cell dimensions and statistical segment lengths.
Finally, the terminal end distribution function, either $q_+(\mathbf{r}, N)$ or $q_-(\mathbf{r}, 0)$, is needed to calculate the chain conformation partition function, which appears as a normalization in the order parameter calculations.
Note that the modular construction of this procedure for calculating the polar order parameter field means that it can be used for arbitrary block compositions and branched architectures, outside the scope of this paper.

\section{\label{app:dithering} Distal bending versus interpenetration}

\begin{figure}[h!]
\centering
\includegraphics[width=2.75in]{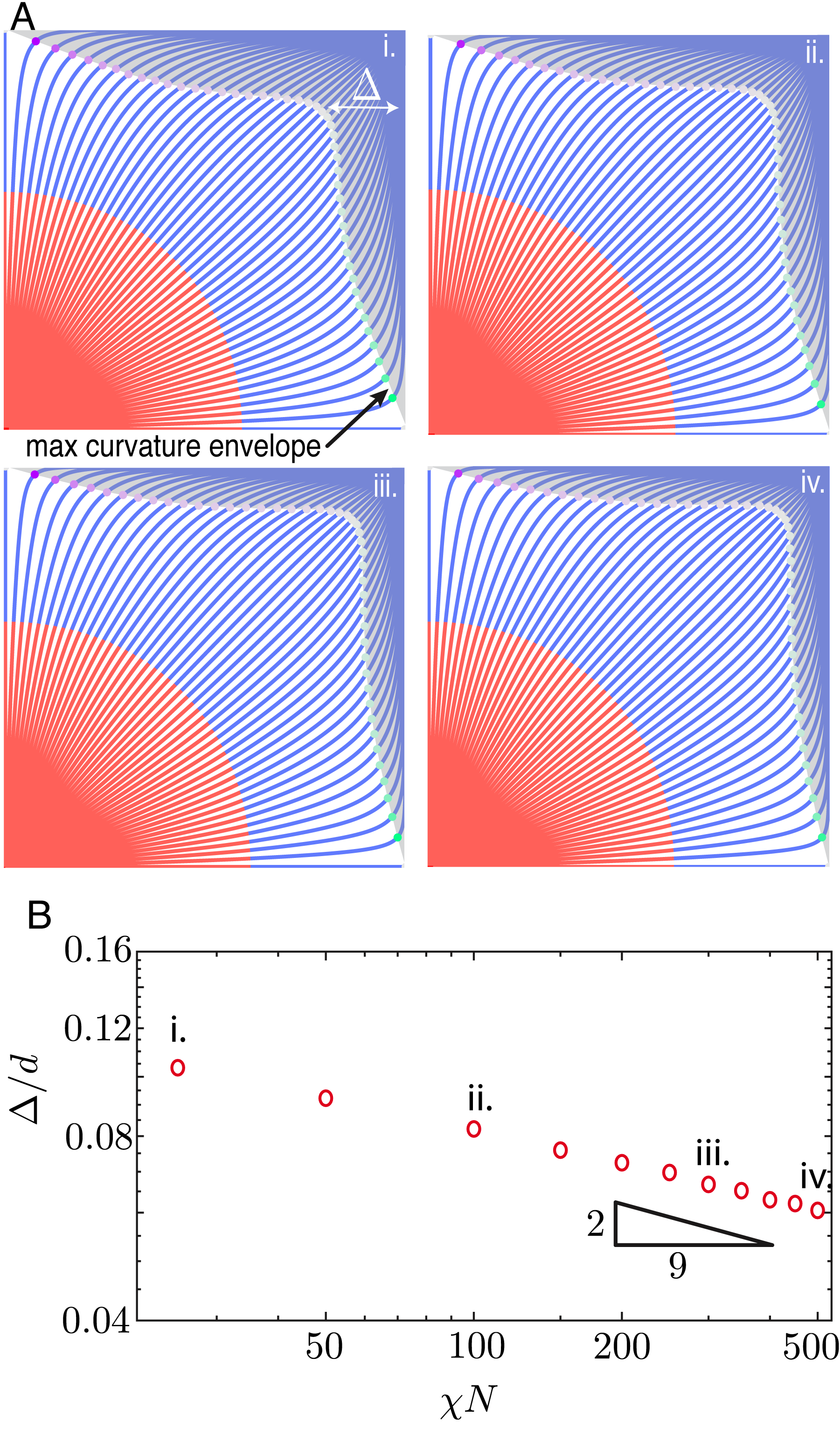}
\caption{\label{fig:dithering} (A) Demarcation of the distal bending region, where interactions between chains from different domains result in pronounced bending, using the maximum curvature of each trajectory. $\Delta$ quantifies the maximum distance of this enveloped from the outer terminal boundary. Examples show the distal bending region for $p4mm$ at (i) $\chi N = 25$, (ii) 100, (iii) 300, and (iv) 500. (B) Log-log plot of the distal bending region width $\Delta$ normalized by unit cell parameter $d$ as a function of $\chi N$, with a $-2/9$ slope indicating the predicted SST scaling the size of the interpenetration zones between contacting brushes~\cite{Goveas1997}.}
\end{figure}

Here, we briefly comment on characteristic deflection of chain trajectories as their distal ends approach outer terminal boundary (see for example Fig.~\ref{fig:local_structure}B).  We note that the magnitude of $\Pv$ tends to zero in this region~\cite{Prasad2017}, as chain trajectories become ``disorientated'' in the contact region between brushes from opposing domains.  Indeed, we observe that the relative size of this ``deflected zone'' compare to the domain sizes decreases with segregation strength, consistent with the size of the interpenetration zone in BCP domains in the $\chi N \to \infty$ limit, predicted to decreases as $(\chi N)^{-2/9}$~\cite{Goveas1997}.  We characterize the edge deflected zone by points of maximal curvature near the distal ends of each trajectory, as highlighted in Fig.~\ref{fig:dithering}A for the $p4mm$ (square) columnar phase.  The decreasing size of this high-bending, distal zone is evident from the sequence of moderate ($\chi N = 25$) to strong ($\chi N = 500$) segregation SCF trajectories.  Fig.~\ref{fig:dithering}B plots the thickness of this distal zone $\Delta$ relative to the domain size as a function of segregation strength, showing it vanishes in proportion to the degree of interpenetration.  Hence, we expect that in $\chi N \to \infty$ limit trajectories tend to a well-defined, limiting configuration that abruptly meets the terminal boundary at finite angle of incidence.

\section{\label{app:bend_threshold} Measures of kink at the IMDS}

\begin{figure}[h!]
\centering
\includegraphics[width=3in]{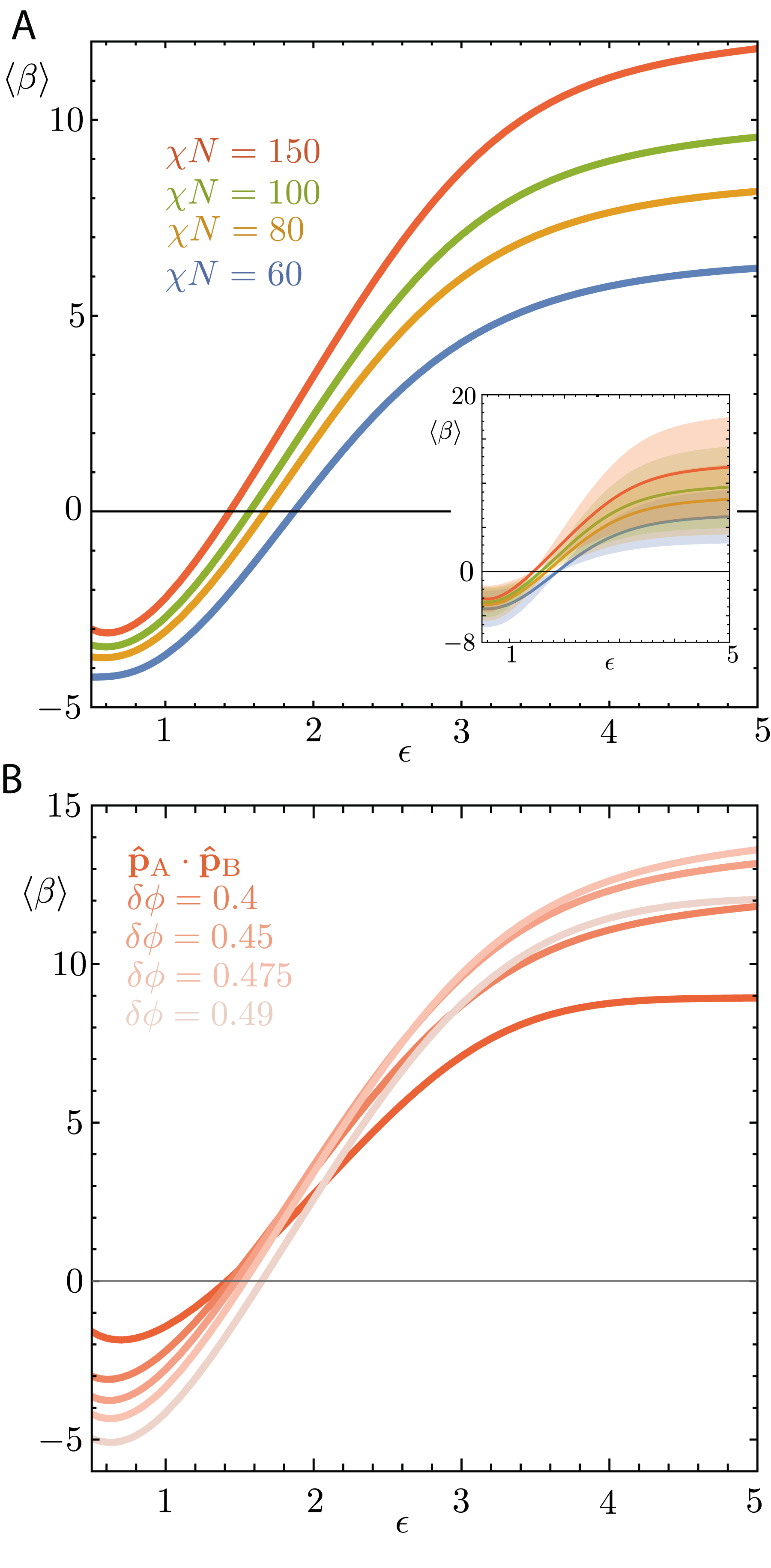}
\caption{\label{fig:bend_threshold} Mean kink angle $\langle \beta \rangle$ calculated from the total polar order parameter $\mathbf{P}$ evaluated on a thresholding window about the IMDS for $p4mm$ for $f = 0.3$. (A) Variation in $\langle \beta \rangle$ calculated at level sets $\phi = 0.1$ and $\phi = 0.9$ ($\delta \phi = 0.4$) with conformational asymmetry $\epsilon$ for $\chi N = 60$ - 150. Inset shows variation in kink angle, with shading representing a single standard deviation about the mean. (B) Variation in $\langle \beta \rangle$ with respect to variation in thresholding window $\delta \phi$ at $\chi N = 150$. For comparison, the at-IMDS kink angle (labeled $\hat{\mathbf{p}}_{\rm A}\cdot \hat{\mathbf{p}}_{\rm B}$) employed in the main text is additionally shown.}
\end{figure}

In the main text, the kink angle $\beta$ is defined through Eq.~(\ref{eq:kink_angle_cos}) as the angle between the A- and B-block polar order parameter fields, respectively $\mathbf{p}_{\rm A}$ and $\mathbf{p}_{\rm B}$, evaluated at the IMDS.
This definition of kink angle is perhaps the simplest and most easily generalizable; it is a {\it strictly local} measure defined at the IMDS.
Alternatively, we can consider a broader estimation of the turning angle of the trajectories as they pass through the IMDS.
To do this, we calculate the orientation of each trajectory $\mathbf{R}_{\mathbf{r}_0}(t)$ at two endpoints $t_-$ and $t_+$ on either side of the IMDS.
Call these two orientations $\hat{\mathbf{P}}_- \equiv \hat{\mathbf{P}}(\mathbf{R}_{\mathbf{r}_0}(t_-))$ and $\hat{\mathbf{P}}_+ \equiv \hat{\mathbf{P}}(\mathbf{R}_{\mathbf{r}_0}(t_+))$.
This alternative kink angle $\beta$ is then taken to be the angle between $\hat{\mathbf{P}}_-$ and $\hat{\mathbf{P}}_+$; like in the main text, averages are then taken over the collection of trajectories with a given fundamental domain.
The two endpoints $t_\pm$ are selected via choice of level sets of the A-block density field $\phi_{\rm A}$.
Since $\phi_{\rm A} = 1/2$ corresponds to the IMDS, we choose $t_\pm$ such that
\begin{equation}
    \phi_{\rm A}(\mathbf{R}_{\mathbf{r}_0}(t_\pm)) = \frac{1}{2} \pm \delta \phi 
\end{equation}
for fixed values of $\delta \phi$.
The resulting kink angle statistics are shown in Fig.~\ref{fig:bend_threshold} for $p4mm$.
As shown in Fig.~\ref{fig:bend_threshold}A, this window-thresholded mean kink angle exhibits similar trends as the at-IMDS kink angle from the main text as a function of $\epsilon$ as well as $\chi N$. 
Moreover, as shown in Fig.~\ref{fig:bend_threshold}B, these two definitions generally approach each other as $\delta \phi$ is decreased (i.e.~the window about the IMDS is taken to be narrower). In other words, as the IMDS window is increased (with the exception of the largest window), the magnitude of kinking generally increases as more of the curved trajectory is taken into account.
Note that in taking $\delta \phi \to 0$, the window-thresholded kink angle will additionally limit to 0 since $\hat{\mathbf{P}}$ is continuous through the IMDS.
However, for non-zero $\delta \phi$, the value of $\hat{\mathbf{P}}_\pm$ will be dominated by the orientation of majority block within a given domain, meaning that to a good approximation, $\hat{\mathbf{P}}_+ \simeq \hat{\mathbf{p}}_{\rm A}$ and $\hat{\mathbf{P}}_- \simeq \hat{\mathbf{p}}_{\rm B}$; this approximation is expected to improve as  $\chi N \to \infty$ since width of the IMDS decreases as $(\chi N)^{-1/2}$~\cite{Semenov1985_SovPhysJETP}.
As $\delta \phi$ is increased maximally ($\delta \phi = 0.49$), the width of the IMDS-window is extended such that the $\hat{\mathbf{P}}$ field becomes increasingly dominated by single-block contributions, the agreement with the at-IMDS definition significantly decreases, which we attribute to ``far-field'' effects of the chain orientation.

% To determine the net turning angle, we express a single trajectory in an arclength parametrization $\mathbf{R}_{\mathbf{r}_0}(s)$, where $s$ is determined such that the trajectory tangents $\partial_s \mathbf{R}_{\mathbf{r}_0}(s)$ are of unit length; per eq.~\ref{eq:int_curves}, this implies that $\partial_s \mathbf{R}_{\mathbf{r}_0}(s) = \hat{\mathbf{P}}(\mathbf{R}_{\mathbf{r}_0}(s))$.
% The bend along each point of the trajectory is then given by $\kappa(s)\hat{\mathbf{n}}(s) = \partial_s \hat{\mathbf{P}}(\mathbf{R}_{\mathbf{r}_0}(s))$, where $\hat{\mathbf{n}}(s)$ is the trajectory normal at arclength parameter $s$.
% The net turning angle accumulated across an arclength interval $s_1 < s < s_2$ is then related to the integral of the bend vector $\kappa(s)\hat{\mathbf{n}}(s)$ over that interval and is given by
% \begin{equation}
%     \left|\int_{s_1}^{s_2} {\rm d}s\, \kappa(s)\hat{\mathbf{n}}(s)\right| = \left|\hat{\mathbf{P}}(s_2) - \hat{\mathbf{P}}(s_1)\right| = 2 \left|\sin\frac{\beta}{2}\right|
% \end{equation}

\section{\label{app:term} Process for locating terminal boundaries}

\begin{figure}[h!]
\centering
\includegraphics[width=3in]{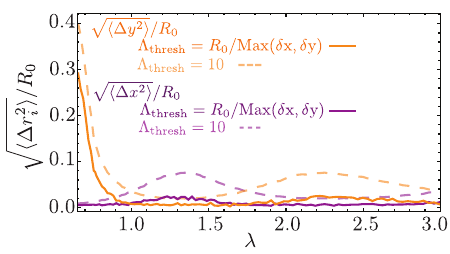}
\caption{\label{fig:lambda_threshold} Variation in the shape of the inner terminal boundary, quantified by the second moments $\sqrt{\langle \Delta x^2\rangle}/R_0$ (purple) and $\sqrt{\langle \Delta y^2\rangle}/R_0$ (orange), for various $c2mm$ structures due to choice of $\Lambda_{\rm thresh}$ as a function of $\lambda$. Solid lines denote the limits of numerical resolution, estimated from the heuristic $\Lambda_{\rm thresh} = R_0/{\rm Max}(\delta x, \delta y) \simeq 30$, whereas dashed lines denote the somewhat relaxed threshold of $\Lambda_{\rm thresh} = 10$ used in Fig.~\ref{fig:c2mm_measure}.}
\end{figure}

To calculate the Jacobian of the association map,  $\Lambda_{ij} (\rv) = \frac{\partial \alpha_i}{\partial r_j}$, we generate a triangular (tetrahedral in 3D) mesh that is refined in the vicinity of the terminal boundaries.
Each mesh facet consists of $d+1$ vertices, where $d$ is the spatial dimension, which we shall label $\mathbf{r}^{(0)}$, $\mathbf{r}^{(1)}$, $\dots$, $\mathbf{r}^{(d)}$.
Using these vertices, we define $d$ edge vectors $\delta \mathbf{r}^{(1)} \equiv \mathbf{r}^{(1)} - \mathbf{r}^{(0)}$, $\dots$, $\delta \mathbf{r}^{(d)} \equiv \mathbf{r}^{(d)} - \mathbf{r}^{(0)}$.
The association map $\bm{\alpha}(\mathbf{r})$ maps these vertices to $d+1$ image points $\bm{\alpha}^{(n)} \equiv \bm{\alpha}(\mathbf{r}^{(n)})$; similarly, we can define $d$ image edge vectors $\delta \bm{\alpha}^{(n)} \equiv \bm{\alpha}^{(n)} - \bm{\alpha}^{(0)}$.
The facet undergoes an affine transformation under the association map, with edge vectors related by the affine matrix $\Lambda$ via
\begin{equation}\label{eq:aff_def}
    \delta \bm{\alpha}^{(n)} = \bm{\Lambda}\,\delta \mathbf{r}^{(n)}\, ,
\end{equation} 
where $\bm{\Lambda}$ takes on a different value for each facet.
Expressing the correction of edge vectors as a row vector $\left[\delta\mathbf{r}^{(1)}\, \dots\, \delta\mathbf{r}^{(d)} \right]$, Eq.~(\ref{eq:aff_def}) can be written as $\left[\delta\bm{\alpha}^{(1)}\, \dots\, \delta\bm{\alpha}^{(d)} \right] = \bm{\Lambda}\,\left[\delta\mathbf{r}^{(1)}\, \dots\, \delta\mathbf{r}^{(d)} \right]$.
This form can then be inverted to solve for $\bm{\Lambda}$,
\begin{equation}
\bm{\Lambda} = \left[\delta\bm{\alpha}^{(1)}\, \dots\, \delta\bm{\alpha}^{(d)} \right]\,\left[\delta\mathbf{r}^{(1)}\, \dots\, \delta\mathbf{r}^{(d)} \right]^{-1} \, ,
\end{equation}
which approximates the Jacobian, becoming exact in the limit $\delta \mathbf{r}^{(n)} \to 0$.

The principal eigenvalue of the Jacobian matrix, $\Lambda$, is used for determining the locations of the terminal boundaries.
We find that the distribution of $\Lambda$ over mesh elements is region-dependent, generally sharply peaked around $\Lambda \simeq 1$, but acquiring a second local maximum near the outer terminal boundary, which we attribute to the outer terminal boundary's role in partitioning space into disconnected domains.
To determine the location of the outer terminal boundary, we choose a threshold value $\Lambda_{\rm thresh}$ to include this second local maximum in the $\Lambda$ distribution; for most computations analyzed here, $\Lambda_{\rm thresh} \simeq 5$ seems to be a reasonable value for demarcation of the outer terminal boundary.
Since the inner terminal boundaries do not partition space into disconnected domains, the distribution of stretching values in the vicinity of these boundaries remain unimodal, requiring a separate heuristic for determining $\Lambda_{\rm thresh}$.
As a first approximation, we estimate calculate the stretching factor that is required to re-scale the SCF grid spacing, $(\delta x, \delta y, \delta z)$, to a characteristic length scale of the IMDS, namely the radius of a sphere with volume equal to the volume $V_{\rm A}$ of the A-block domain (or area $A_{\rm A}$ in 2D), $R_0 = \big(3 V_{\rm A}/(4\pi)\big)^{1/3}$ (or $R_0 = \big(A_{\rm A}/\pi \big)^{1/2}$), so $\Lambda_{\rm thresh} \simeq R_0/{\rm Max}(\delta x, \delta y, \delta z)$. 
Using this heuristic, we find $\Lambda_{\rm thresh} \sim \mathcal{O}(10)$, with typical values of 10 - 30 for the inner terminal boundary, depending on the geometry of the IMDS and the spatial grid resolution of the SCF calculation.
For highly non-spherical IMDSs, we find that smaller values of $\Lambda_{\rm thresh}$ are required to determine finder features of the inner termini.
In regards to our heuristic, this is because the IMDS develops finer-scale features, such as variable radii of curvature that are $\lesssim R_0$, leading to a depressed $\Lambda_{\rm thresh}$.
Moreover, since the heuristic for $\Lambda_{\rm thresh}$ estimates the limit based on numerical resolution, the resulting terminal boundaries are sensitive to discretization error.
As a result, for the $c2mm$ calculations, we use a lower threshold, choosing $\Lambda_{\rm thresh} \simeq 10$ rather than $\simeq 30$ based on the heuristic.
As shown in Fig.~\ref{fig:lambda_threshold}, the shape and size of the $c2mm$ inner terminal boundary depends on the choice of $\Lambda_{\rm thresh}$, but importantly the qualitative features are retained even at the limits of numerical resolution.

\bibliography{refs}

\end{document}